# Causal Discovery via Simultaneous DAG Recovery Using the Angles Space of Directional Dependence Measures

## JD Opdyke, Chief Analytics Officer, DataMineit, LLC

**Abstract**

Most causal discovery algorithms utilizing a Directed Acyclic Graph (DAG) framework execute sequentially (e.g. constraint-based models) or iteratively (e.g. score-based or functional causal models). In contrast, we develop a new method – Angles-based Directional Dependence (ADD) – that executes over the entire DAG space simultaneously, based on only two matrix estimations. We apply dual orderings on any (positive definite) directional dependence measure for all pairwise relationships, identify statistically significant directional dependence in the (positive definite) angles space, and then enforce acyclicity to make proper causal interpretations. Potential benefits of the approach include increased coherence, due to simultaneous edge-calling within a positive definite space; increased power, due to the ability to use any directional dependence measure and thus, opportunistically adapt to different or varying data conditions; and increased speed/scalability, due to the need for only two matrix estimations, and two (fast) simulations to define empirical confidence bounds under independence (regardless of the size of the DAG space). We conduct a preliminary empirical study evaluating direct adjacency under nonlinear, asymmetric, heavy-tailed data conditions. The promising results indicate applications to feature selection in quantitative finance, and justify and encourage a more extensive follow-up study to benchmark against competing algorithms to more fully test the above-mentioned potential benefits of ADD.

**Keywords**: directional dependence, causality, causal discovery, DAG, directed acyclic graph, causal modeling, quantitative finance, positive definite, feature selection, dependence structure, Pearson's, covariance, correlation, Chatterjee's, direct adjacency

**Acknowledgments:** I thank Alejandro Dominguez Rodriguez, PhD, for his very insightful comments and useful suggestions on earlier drafts of this paper.

**Contents**

## 1. Introduction

Directed Acyclic Graphs (DAGs) are a widely used and well-established framework for causal modeling,[1] which includes both causal discovery and causal inference (see Zanga et al., 2025). The former identifies causal relationships and the latter quantifies them. This paper focuses only on causal discovery, seeking to identify causal structure consistent with the DAG representing the 'true' network of causal relationships. Many algorithms designed to recover DAG 'truth' execute sequentially. These include constraint-based methods, such as PC algorithms (see Kalisch, M., & Bühlmann, P. (2007)), FCI algorithms (see Sprites (2001)), and Markov Blanket algorithms (see Aliferis et al. (2010), and Chen et al., (2026)). Most of the other approaches execute iteratively. These include score-based methods, such as NOTEARS (see Zheng et al. (2018)) and its variants (e.g. DYNOTEARS, see Pamfil et al. (2020)), Greedy Equivalence Search (see Chickering, D., (2002) and Hauser and Bühlmann (2012)) and Kernel-based Score Functions (see Huang et al., (2018) and Wang et al., (2024)). They also include functional causal models, such as LiNGAM (see Shimizu et al. (2006)), and its variants such as VarLiNGAM (see Hyvärinen et al. (2010)). In contrast, the method developed herein – Angles-based Directional Dependence (ADD) – estimates the entire DAG structure simultaneously, via dual orderings using any positive definite directional dependence measure. So ADD requires only two estimations of the all-pairwise dependence matrix, one in each direction to establish directionality, and two simulations to establish empirical confidence bounds, under independence, for each cell of the matrix. Inference takes place in the angles space (based on the empirical angles distributions generated from the simulations).[2] Potential benefits of this approach include increased coherence, due to simultaneous edge-calling within a positive definite space;[3] increased power, due to the ability to use any (positive definite) directional dependence measure and thus, opportunistically adapt to different or varying data conditions; and increased speed/scalability, due to the need for only two matrix estimations and two (fast) simulations (regardless of the size of the DAG space). This describes ADD's recovery of directional dependence structure, but making a valid causal interpretation requires satisfying additional assumptions specified below (see A1-A7 in Appendix C).[4]

[1] Of course, DAGs are not the only framework for causal discovery. The potential outcomes framework (see Angrist et al. (1996) and Imbens and Rubin (2015)) is the major competitor to Structural Causal Models (the overarching framework utilizing DAGs). Czado (2025) also demonstrates that vine copulas can be remarkably effective in the causal discovery setting. And Rodriguez Dominguez & Yadav (2024) and Rodriguez Dominguez (2026) propose approaches to causal measurement based on system-level and spectral properties, respectively. We address potential and actual limitations of DAGs when used for causal discovery in later sections herein.

[2] Spherical angle parameterizations of dependence measures go back at least to Fisher (1915). See Joarder and Ali (1992), Rapisarda et al. (2007), and Zhang et al. (2015) for detailed derivations and geometric interpretations of their construction.

[3] The "wobbly world" of causal graphs, often characterized by severe instability and lack of coherence in edge-calling across a DAG, is convincingly tested and demonstrated in Hulse et al. (2025). Similarly, Faltenbacher et al. (2026) carefully document the sources and magnitude of incoherent errors in DAG-based causal discovery. Finally, Padh et al. (2025) describe how a popular class of DAG recovery algorithms (i.e. those that are 'score-based') critically depend on heuristic hyperparameter choices and remain sensitive to threshold choices.

We first review the long and more recent history of directional dependence measures, and their more recent use and utility for causal modeling. Next, we do the same for the use of the angles space in measuring and estimating dependence in a wide range of settings, including quantitative finance. Finally, we combine the two to develop ADD, which utilizes the (positive definite) angles space of (positive definite) directional dependence measures to identify causal relationships within the DAG framework. We conduct a preliminary empirical study to test ADD using conditioning to obtain valid results under direct adjacency. We use highly nonlinear, asymmetric, and heavy-tailed data generated by a straightforward, but non-toy DAG and obtain promising results, indicating applications to feature selection in quantitative finance and justifying further inquiry. The latter takes the form of a proposed, more extensive study explicitly benchmarking ADD against causal discovery competitors to more fully test its proposed benefits under wide-ranging conditions.

## 2. Background

### 2.A. Asymmetric, Directional Dependence

We present a brief survey of the relevant literature on directional dependence here, with a focus on readily implemented methods (covering (1) through (6), and (10) through (13) below), because these are central to the development of ADD and have been extensively tested empirically within the ADD framework. The concept of asymmetric, directional dependence is not new. Recent research developing such measures goes back over a dozen years (see Zheng et al., 2012), but has its direct origins in work done at the end of the nineteenth century (see Yule, 1897, and a related review in Allena and McAleerb (2018)). As discussed in more detail below in the section covering the angles space, ADD's reliance on the Cholesky factor requires the dependence measures it uses to be positive definite. This is a defining feature of dependence measures generally, so for clarity throughout this paper, and with practical, empirical settings in mind, we group dependence measures vis-à-vis positive definiteness into three mutually exclusive and exhaustive categories:

C1. Measures that have been analytically proven to be positive definite, and which only exhibit non-positive definiteness empirically under select conditions due to numerical issues. Examples include Pearson's rho, Spearman's rho, and Kendall's tau (see Sabato (2007) for corresponding proofs) as well as the tail dependence matrix (see Embrechts (2016) for a corresponding proof). Note that none of these measures are directional.

C2. Measures for which positive definiteness has not been proven analytically, but which, via extensive empirical simulations and testing under wide-ranging conditions, do not violate positive definiteness any more frequently than those measures from category C1., thus indicating that such violations are likely due only to numerical issues. Examples include the directional measures (1) through (6), and (10) through (13) presented below.

[4] Briefly, these include acyclicity, causal sufficiency, Markov and faithfulness, independent and identically distributed sampling, reliable measurement, and monotone link from (directional) dependence measure to asymmetry.

C3. Measures for which positive definiteness has not been proven analytically, and which exhibit non-positive definiteness more often than numerical issues alone would indicate.

The directional dependence measures reviewed below, used with ADD, and used in the empirical study herein all fall into category C2. While empirical findings should never be taken as evidence on the same level as analytical proof, the development and testing of ADD, using all eight of the category C2. measures, required many millions of simulations under wide-ranging data conditions, and these never once yielded a single non-positive definite result: hence their classification in category C2. Best practices dictate that positive definiteness always should be empirically verified anyway, even for C1. measures, due to the possibility of numerical issues, especially when the factor space is not small (spaces with #factors = 100 were tested), so we do not view the lack of analytic proofs for C2. measures as posing a limiting restriction on their use for these purposes in practice. If the rare edge case yields a non-positive definite matrix due to numerical issues, one can use a 'fixing' algorithm whereby the dependence measure matrix is projected to the nearest positive definite matrix (see the seminal work of Higham (2002) for a widely used example).

Category C3. measures, on the other hand, warrant more caution. While one can always mechanically turn to algorithms like Higham (2002) to enforce positive definiteness, if this occurs due to the construction of the measure and not due solely to numerical issues when performing sampling, this approach would not be advisable. For example, if we observed non-positive definiteness in more than, say, one or two percent of samples (admittedly a subjective threshold), we would avoid this 'matrix-fix' approach (and likely avoid the dependence measure altogether).[5] The main problem is that such an approach would materially distort the sample space of the dependence measure, making inferences based on such an automated, but statistically invalid ruleset potentially highly misleading, in *either* direction: for example, by increasing false negatives by missing true signals, or by increasing false positives by wrongly flagging nonexistent signals, or even both by increasing both simultaneously. The bottom line here is that ADD requires the dependence measures it uses to be positive definite, and implementation of those that fall into category C2., all of which have been tested extensively by ADD, is defensible for practical, scientifically vigilant application in this setting. We review these eight measures below.

---

[5] To flip this script on this requirement of positive definiteness, one reasonably could argue that if a dependence measure was analytically shown to be non-positive definite, at least under relevant conditions, and/or its empirical estimates were non-positive definite more often than could be attributable solely to numeric considerations, then researchers and practitioners might want to question the wisdom of using it. Non-positive definiteness also could be a function of unknown or mis-specified data conditions, such as perfect linear dependence unwittingly built into a simulation (although with actual market data, it could be a very useful flag for extreme multicollinearity). Or non-positive definiteness perhaps could be due to a combination of the dependence measure used and the specific data conditions being examined. Either way, the non-positive definite results could be serving as a correct and useful warning to avoid the dependence measure (and/or those simulated data conditions) altogether. In such cases, the requirement of positive definiteness is less a limitation of a method like ADD and more a proper boundary on the right measures and conditions under which such analyses should be conducted in the first place.

A recent example of a directional dependence measure that garnered much attention upon its publication in 2021 is Chatterjee's new correlation coefficient (chcorr). This is largely due to its simplicity and ease of implementation as a measure of non-linear, non-monotonic, regression-based, and cyclical dependence. If observation pairs of X and Y variables are ranked according to X values, with no ties on the X values, so that $\left(\left(X_{(1)},Y_{(1)}\right),\left(X_{(2)},Y_{(2)}\right),\cdots,\left(X_{(n)},Y_{(n)}\right)\right)$ then:

(1) $$chcorr = \xi_n\left(X,Y\right) := 1 - \frac{3\sum_{i=1}^{n-1}\left|r_{i+1}-r_i\right|}{n^2-1} \text{ where } r_i = \text{rank of } Y_i$$

Under ties for some of the X values, break ties uniformly at random, and

(2) $$chcorr = \xi_n\left(X,Y\right) := 1 - \frac{n\sum_{i=1}^{n-1}\left|r_{i+1}-r_i\right|}{2\sum_{i=1}^{n} l_i\left(n-l_i\right)} \text{ where } l_i = \#j \text{ such that } Y_{(j)} \geq Y_{(i)}$$

Chatterjee's new correlation coefficient ranges from zero to one asymptotically (it can exceed these bounds slightly under finite samples), and unlike many commonly used dependence measures, a value of zero indicates independence. Also unlike some other measures, no positive or negative dependence is indicated by a positive or negative sign on the measure value.[6] As an asymmetric dependence measure, the order of X and Y matters: $\xi_n\left(X,Y\right)$ does not necessarily equal $\xi_n\left(Y,X\right)$, by design. In other words, the dependence of Y on X is not assumed to be identical to the dependence of X on Y, respectively: dependence is *directional*.[7] However, note that Chatterjee's can be made to be symmetric (at the pairwise cell level) by simply taking the maximum of two measures, one in each direction as in (3):

(3) $$chcorr_sym = \max\left[\xi_n\left(X,Y\right),\xi_n\left(Y,X\right)\right]$$

Chatterjee's breakthrough has spawned many variants (see Lin & Han (2023), Pascual-Marqui et al. (2024), and especially Gao and Li (2024)). One of these is the "improved Chatterjee's correlation"

---

[6] Recall, of course, that widely used non-directional dependence measures such as Pearson's rho, Spearman's rho, and Kendall's tau have (maximum) ranges of –1 to 1, indicate positive/negative dependence with their signs, and values of zero do not necessarily indicate independence. A counter example of the latter, however, is Pearson's under Gaussian data, whereby a value of zero *does* indicate independence.

[7] This holds true for all asymmetric/directional dependence measures, so importantly, the corresponding all-pairwise matrix is forced to be symmetric. In other words, when using, say, Chatterjee (2021) on two factors, say, X3 and X4, the value in cell row 3, column 4 of the matrix is $\xi_n\left(X3,X4\right)$, and the value in cell row 4, column 3 of the matrix is identical, that is, $\xi_n\left(X3,X4\right)$; it is NOT $\xi_n\left(X4,X3\right)$. Stated differently, the matrix can be thought of initially as a lower triangular matrix, whose values are subsequently placed in the upper triangle (along with unit diagonals) to make the matrix symmetric. To obtain a symmetric matrix measuring directional dependence of all the pairs in the other direction, as described below, the data columns are simply reversed, a new lower triangle of values is obtained and those values are placed in the upper triangle.

(ichcorr, or "ICH") derived by Xia et al. (2025)), the motivation of which is to increase power by using inverse distance weightings of all neighboring data values as opposed to just one.

(4) $$ichcorr = \xi_n^{IM}\left(X,Y\right) = 1 - \frac{\sum_{i \neq j}^{n} \left|r_i - r_j\right| / \left|i - j\right|}{\frac{n+1}{3}\sum_{i \neq j}^{n} 1 / \left|i - j\right|}$$

Xia et a. (2025) test the power and level of "improved Chatterjee" against other dependence measures, including Chatterjee's, in an empirical study under wide-ranging data conditions. Both Chatterjee's and improved Chatterjee's coefficients exhibit good power under non-monotonic, non-linear, and cyclical dependence, with the latter measure usually winning. Like Chatterjee's, ICH ranges from zero to one asymptotically,[8] has no sign indicating positive or negative dependence, and indicates independence with a value of zero.[9]

Interestingly, Zhang (2024a) has proposed combining Chatterjee's and Spearman's measures as in (5) in an effort to obtain the best of both worlds: a dependence measure that has reasonable power under cases of non-monotonic, non-linear, and/or cyclical dependence (where Spearman's has little to no power, especially compared to Chatterjee's) as well as reasonable power under monotonic dependence (where Chatterjee's has less power than Spearman's).

(5) $$zcorrsp = I_{n_sp}\left(X,Y\right) = \max\left\{\left|sr_{X,Y}\right|, \sqrt{5/2}\,\xi_n\left(X,Y\right)\right\}$$

Zhang's (2024a) combined correlation is signless, and zero indicates independence. This dependence measure also is directional/asymmetric due to its inclusion of Chatterjee's coefficient. Zhang (2024b) later derived the symmetric version of this test as (6):

(6) $$zcorrsp_sym = \max\left\{\left|sr_{X,Y}\right|, \sqrt{5/2}\,\xi_n\left(X,Y\right), \sqrt{5/2}\,\xi_n\left(Y,X\right)\right\}$$

For completeness, Spearman's rho is defined below:

(7) $$sr_{X,Y} = \frac{\sum_{i=1}^{n}\left(R_{X_i} - \frac{1}{n}\sum_{j=1}^{n} R_{X_j}\right)\left(R_{Y_i} - \frac{1}{n}\sum_{j=1}^{n} R_{Y_j}\right) \Big/ \left(n-1\right)}{\sqrt{\sum_{i=1}^{n}\left(R_{X_i} - \frac{1}{n}\sum_{j=1}^{n} R_{X_j}\right)^2 \Big/ \left(n-1\right)}\sqrt{\sum_{i=1}^{n}\left(R_{Y_i} - \frac{1}{n}\sum_{j=1}^{n} R_{Y_j}\right)^2 \Big/ \left(n-1\right)}}$$

where R indicates the rank ordering of the subscripted variable. In the event of ties in the ranks,

[8] Note that like Chatterjee's correlation, the finite sample distribution of 'improved' Chatterjee's correlation can exceed these bounds.

[9] See Appendix B for code implementing ICH that efficiently increases speed by a factor of p, the dimension of the matrix.

widespread convention is to use the averaged rank (see Zar, 1999). If there are no ties in the data, (7) can be shortened to

(8) $sr_{X,Y} = 1 - \frac{6\sum_{i=1}^{n}\left(R_{X_i} - R_{Y_i}\right)^2}{n^3 - n}$, although this version also can be adjusted for ties as in (9):

(9) $sr_{X,Y} = \frac{\left(\left[n^3-n\right]/6\right) - \sum_{i=1}^{n}\left(R_{X_i}-R_{Y_i}\right)^2 - \sum_{i=1}^{mx}\left(t_{X_i}^3 - t_{X_i}\right)/12 - \sum_{i=1}^{my}\left(t_{Y_i}^3 - t_{Y_i}\right)/12}{\sqrt{\left[\left(\left[n^3-n\right]/6\right) - \sum_{i=1}^{mx}\left(t_{X_i}^3 - t_{X_i}\right)/6\right]\cdot\left[\left(\left[n^3-n\right]/6\right) - \sum_{i=1}^{my}\left(t_{Y_i}^3 - t_{Y_i}\right)/6\right]}}$ where mx and my=# tied groups in x and y, respectively, and $t_{X_i}$ and $t_{Y_i}$ are the number of ties in each tied group.[10]

Another measure similar to Chatterjee's is the Differential Distance Correlation (DDC) of Liu and Shang (2025). DDC's values also are signless and range from zero to one, with zero indicating independence. Notably, similar to the widely used Szekely's distance correlation (see Szekely et al. (2007)), DDC can be multidimensional, but when X is univariate so that DDC can be used in an all-pairwise matrix, it is defined as (10) below:

(10) $DDC_n\left(X \mid Y\right) = 1 - \frac{1}{(n-1)}\sum_{i=1}^{n-1}\left\|X_{(i)} - X_{(i+1)}\right\| \Big/ \left[\binom{n}{2}^{-1}\sum_{i=1}^{n}(2i-n-1)X^{(i)}\right]$

where $\left\{\left(X_{(i)}, Y_{(i)}\right)\right\}_{i=1}^{n}$ are ordered to satisfy $Y_{(i)} \le \cdots \le Y_{(n)}$, and

$X^{(i)}$ are ordered to satisfy $X^{(i)} \le \cdots \le X^{(n)}$. Liu and Shang (2025) show in an empirical study that DDC has power similar to that of Chatterjee's measure, but with slightly more power under damped oscillator data. And like Chatterjee's measure, DDC is directional, so in the general case, $DDC_n\left(X \mid Y\right) \ne DDC_n\left(Y \mid X\right)$. Also like Chatterjee's measure, a symmetric, non-directional version may be obtained via (11)[11]:

(11) $DDC_{n-sym}\left(X,Y\right) = \max\left[DDC_n\left(X \mid Y\right), DDC_n\left(Y \mid X\right)\right]$

Yet another recently developed measure based on the original Chatterjee (2021) correlation is derived in Yin (2026) as the 'modified' Chatterjee correlation (mch), with finite-sample estimator (for continuous variables) shown in (12):

---

[10] Averaging the ranks of ties in (7) is computationally preferred to (9) (see Zar (1999)).

[11] Based on email correspondence with author Yixiao Liu, July 9, 2025.

(12) $mch = \xi_{+,n}(X,Y) = n\sum_{i=1}^{n}\sum_{j=1}^{n}\left(\frac{R_{i,j}}{n_j} - \frac{R_i}{n}\right)^2 \Big/ \sum_{i=1}^{n} R_i\left(n - R_i\right)$ where

$R_i = \sum_{k=1}^{n} 1_{\{Y_k < Y_i\}}$, $n_j = \sum_{k=1}^{n} K_h\left(X_k - X_j\right)$, $R_{i,j} = \sum_{k=1}^{n} 1_{\{Y_k < Y_i\}} K_h\left(X_k - X_j\right)$, and $K_h(t) = K\left(\frac{t}{h}\right)\Big/h$ is the Gaussian kernel with bandwidth = *h* (the simulation study uses Silverman's (1986) bandwidth of $h = 1.06\hat{\sigma}_X n^{-1/5}$ ). This carefully derived measure eliminates some of the arguable shortcomings of the other Chatterjee-related measures, such as finite sample distributions whose values extend below zero.

Finally, asymmetric, directional dependence measures also can be applied only to the tails of X and Y, and for financial settings, it is important to note that correlation breakdowns often are associated specifically with (asymmetric) tail dependence, as described by Pramanik (2024): "Extensive evidence has been gathered showcasing the prevalence of heavy-tailed distributions and asymmetric tail interdependence within equity and foreign exchange markets, particularly during times of crisis." And from Ito and Yoshiba (2025): "We provide new evidence that lower tail dependence coefficients increased compared to upper ones for all pairs in the COVID-19 crash..." One straightforward example of an asymmetric tail dependence measure is that of Deidda et al. (2023) which is essentially Kendall's tau applied conditionally, only when the percentile, *q*, of X (or Y) is exceeded:

(13) $\hat{\tau}_{X,Y}(q) = \left(2(k-2)!/k!\right) \sum_{1 \le i \le j \le n} \text{sgn}\left(X_i - X_j\right)\text{sgn}\left(Y_i - Y_j\right) I\left(X_i, X_j > X_{(n-k)}\right)$

where $q = 1 - k/n$, and $2 \le k < n$ is the number of exceedances used in the tail, and $I(\ )$ is the indicator function (one when true, zero otherwise) ensuring that only the *k* largest observations of X are used. Note again that generally, $\hat{\tau}_{X,Y}(q) \ne \hat{\tau}_{Y,X}(q)$, that is, this tail dependence measure is directional, and the effect of X's tail on Y's tail is not assumed to be the same as that of Y's tail on X's tail.

Other directional, asymmetric dependence measures include the dynamic asymmetric tail dependence measure of Ito and Yoshiba (2025), the QAD measure of Junker et al. (2021), the generalized correlation of Zheng et al. (2012), the extremal directional dependence measure of Garcin and Nicolas (2026), the extreme risk network approach of Klüppelberg and Krali (2026), the multivariate directional tail-weighted dependence measure of Li and Joe (2024), and the measures of Vinod (2022) and others described in Jondeau (2016).[12]

[12] We cite these measures for completeness, but do not focus on their implementation in this setting as they are more computationally involved and less readily implemented compared to (1) through (6), (10) through (13). On a related note, Metsämuuronen (2022) intriguingly demonstrates that under certain conditions, such as when categorical and ordinal data are being analyzed and the number of categories between the two variables differs dramatically, even Pearson's correlation can be unambiguously directional.

### 2.B. Relationship between Measures of Association and Causality

This brief review of some of the more widely used and readily implemented directional dependence measures is relevant here because of their usability and utility within causal models, including ADD. But this utility holds not only for directional measures of association, but also for some of the oldest and most widely used non-directional measures of association, such as the covariance matrix. For example, Rodriguez Dominguez (2023, 2025a), together with subsequent work on causal PDE-control models for dynamic portfolio optimization (see Rodriguez Dominguez (2025c)), demonstrate that association-based structures, such as covariance and sensitivity measures, can be embedded within causal frameworks through structured representations of common drivers. Similarly, Cai et al. (2025) utilize the Cholesky factorization of the covariance matrix as the foundation of their causal algorithm, and this turns out to be a not uncommon approach (see also, for example, Li et al. (2025)). Additionally, Pascual-Marqui et al. (2024) ingeniously combine Chatterjee's and Szekely's measures to effectively perform directional, causal regressions. The extant literature contains many similar examples (see also Blömbaum et al. (2019) and MacKinnon & Lamp (2021)), making the point that some of the best applied causal modeling frameworks are those that in no small part intelligently utilize the most widely used and well-established association-based dependence measures,[13] especially those that are asymmetric / directional.

This should come as no surprise because, as noted above, while the development of directional dependence measures is not new, going back to the late nineteenth century (see Yule (1897)), neither is causal modeling, which goes back at least to Wright (1921). While gratuitous resistance to more recent and seminal work on causality (see Pearl (2000, 2009, and 2018)) undoubtedly plagued its introduction, we caution against the other end of the spectrum, that is, portraying causal modeling as brand new, separate from, and/or superior to association-based modeling.[14] This can provide a convenient strawman for argumentative debates in the form of the tired mantra "correlation is not causation," but for rigorous, applied research it draws a false, bright line that seriously limits the advancement of causal modeling, precluding even from preliminary consideration the development of some of the most effective current (and future) approaches![15]

Making strides toward the effective application of causal frameworks, especially in quantitative finance, will be incremental and based in nontrivial ways on over a century of rigorous association-based research and methodology development, even in the face of paradigm shifts; claims to the contrary should be met with appropriate scientific skepticism. For a compelling empirical study refuting some of the overly strong promotional claims made regarding the application of causal models to, for example, quantitative finance, see Rodriguez Dominguez (2025b). This work convincingly demonstrates that while 'proof' of causality is desirable, it is not per se *necessary* for many objectives, such as portfolio optimization.

---

[13] Consistent with this theme, note also that the selection of causal drivers in Rodriguez Dominguez (2023, 2025a) follows the Reichenbach Common Cause Principle (Reichenbach, 1956), a foundational idea in probabilistic causality which emphasizes the identification of common causes through observed ***correlations***.

[14] As Moodie and Stephens (2022) find in their review, "…it is false … to suggest that ordinary, conditional regression models *cannot* provide causal estimates: although there are some settings in which this is true, it is not universally the case."

[15] For example, one book promoting the use of causal models in quantitative finance mistakenly characterizes association-based dependence, which covers an enormous swath of all statistical work in the past century, as being strictly non-directional: "Third, unlike association, causality is directional" p.5, Lopez de Prado (2023). Such oversights ignore the long-established, rich, and useful literature on (association-based) directional dependence measures described above, thus limiting innovation in applied causal research.

So rather than making overreaching and/or unsubstantiated claims regarding causality *versus* association-based methods, it is far more accurate and appropriate and *useful* to say that measures of association, both directional and non-directional, often serve as effective foundational supports for causal methods. Correlation may not be causation, and in multivariate settings the longtime challenge has been, and continues to be, avoiding the attribution of causality to relationships where only spurious association exists. However, correlation and other measures of association undoubtedly are often very strong *indicators* of causation, and the responsible use of association-based dependence measures elevates and underpins some of the best causal models.[16] Ignoring over a century of rigorous development of association-based dependence measures when tackling difficult causal problems is, in fact, the limiting path, throwing the proverbial baby out with the bathwater. Causal modelers should not fear or eschew association-based measures of dependence, but rather, carefully and responsibly and rigorously embrace them.[17] This is the path ADD takes to accurate and powerful causal discovery and DAG recovery, as described below.

### 2.C. The Angles Space

To describe the development of ADD, we must first review in this section the angles space and the related literature, and then in the next section, its specific use with directional dependence measures. Reliance on spherical angles and hypersphere parameterizations is increasingly common in quantitative finance (see for some examples Golts & Jones, 2009; Li, Q., 2018; Hellton, 2020; Zhang, 2022; and Saxena et al., 2023), in large part due to its scale invariance. It has even been used to define entire financial markets (see Kim and Lee, 2016),[18] as well as in other settings to solve, with greater robustness and efficiency, complex multivariate problems (see Zhang and Yang (2025)). This more recent utilization is based on a long history, going back to the origin of modern statistics. The use of spherical angles for the analysis specifically of Pearson's correlation matrix goes back at least to Fisher (1915, 1928), but Joarder & Ali (1992) and Rapisarda et al. (2007) provide geometrically motivated, thorough, and clear descriptions of its derivations. The bivariate version of this is simply the widely known "cosine similarity" shown in (14) below, where Pearson's correlation is defined by the cosine of the angle between two mean-centered variables, which equals the inner product divided by the product of the two vectors' (Euclidean) norms.

---

[16] Per Moodie and Stephens' (2022) review of the causal methodology literature: "... we review core tenets and methods of causal inference ... highlight connections with traditional 'associational' statistical methods ... and point to current topics of active research in this crucial area of our field."

[17] As Moodie and Stephens (2022) conclude, "Causal inference should not be seen as conflicting with traditional statistical methods of explanatory analysis, and causal thinking should be considered relevant even in predictive modeling ..."

[18] Note that even for purposes of <u>*estimating*</u> dependence measure values, spherical angles can be among the best approaches. As determined in Pinheiro and Bates (1996): "Of the five parameterizations considered here, the spherical parameterization presents the best combination of performance and statistical interpretability of individual parameters."

(14) $$\cos\left(\hat{\theta}\right)=\frac{\text{inner product}}{\text{product of norms}}=\frac{\langle \mathbf{X},\mathbf{Y}\rangle}{\|\mathbf{X}\|\|\mathbf{Y}\|}=\frac{\sum_{i=1}^{n}\left(X_i-\frac{1}{n}\sum_{j=1}^{n}X_j\right)\left(Y_i-\frac{1}{n}\sum_{j=1}^{n}Y_j\right)}{\sqrt{\sum_{i=1}^{n}\left(X_i-\frac{1}{n}\sum_{j=1}^{n}X_j\right)^2}\sqrt{\sum_{i=1}^{n}\left(Y_i-\frac{1}{n}\sum_{j=1}^{n}Y_j\right)^2}}=\frac{\hat{Cov}_{X,Y}}{s_X s_Y}=r_{X,Y},\ 0\le\hat{\theta}\le\pi$$

It turns out that the matrix analogue for cosine similarity is well established in the literature (see Joarder and Ali (1992), Pinheiro and Bates (1996), Rapisarda et al. (2007), Pourahmadi and Wang (2015), and Cordoba et al. (2018)). Translation between the (positive definite) dependence measure matrix and the angles matrix is shown below, both formulaically in (15)-(17) and in SAS/IML computer code in Table A. This is, in fact, a multivariate bijection, so each (positive definite) dependence measure matrix corresponds uniquely, one-to-one, to a matrix of angles. The steps for translating between dependence measure values and angles, in both directions, are shown in A.-C. below.

A. estimate the correlation (dependence measure) matrix from sample data
B. obtain the Cholesky factorization of this matrix
C. apply inverse trigonometric and trigonometric functions to B. to obtain the corresponding matrix of spherical angles

and in reverse:

C. start with a matrix of spherical angles
B. apply trigonometric functions to C. to obtain the Cholesky factorization
A. multiply B. by its transpose to obtain the corresponding matrix of the dependence measure
(see Rapisarda et al. (2007), and Pourahmadi & Wang (2015))

Central to this translation mechanism is obtaining the Cholesky factor of the dependence measure matrix, which is usually a built-in function in most statistical and mathematical software. The relevant formulae are included below in (15) for completeness.

(15) A (dependence measure) matrix *R* will be real, symmetric positive-definite,[19] so the unique matrix *B* that satisfies $R=BB^T$ where *B* is a lower triangular matrix (with real and positive diagonal entries), and $B^T$ is its transpose, is the Cholesky factorization of *R*. Formulaically, *B*'s entries are as follows:

$$B_{j,j}=\sqrt{R_{j,j}-\sum_{k=1}^{j-1}B_{j,k}^2}\ ,\qquad B_{i,j}=\frac{1}{B_{j,j}}\left(R_{i,j}-\sum_{k=1}^{j-1}B_{i,k}B_{j,k}\right)\ \text{for } i>j$$

[19] Positive semi-definiteness includes the case of eigenvalues exactly equal to zero, which we largely ignore herein as a border case relevant mainly for textbook examples, since variables (whether financial returns, or factors) would have to exhibit exact linear dependence for an eigenvalue to be exactly zero.

The Cholesky factor, which is only defined under positive definiteness, can be viewed as a matrix analog to the square root of a scalar, because similar to a square root, the product of it and its transpose yields the original matrix. Importantly, the Cholesky factor places us on the UNIT hyper-(hemi)sphere (where scale does not matter) because the sum of the squares of its rows always equals one. Next, we recursively apply inverse trigonometric and trigonometric functions to each cell of the Cholesky factor to obtain each cell's angle value per (16); or in reverse, we obtain a dependence matrix value from trigonometric functions applied to each cell's angle value per (17) (see both Joarder & Ali (1992) and Rapisarda et al. (2007) for meticulous derivations of these formulas). Again, note that this relationship is a multivariate bijection: a one-to-one, invertible, strictly monotone correspondence,[20] with a unique dependence measure matrix yielding a unique angles matrix, and vice versa.

(16) For $R$, a p x p correlation matrix, $R = \begin{bmatrix} 1 & r_{1,2} & r_{1,3} & \cdots & r_{1,p} \\ r_{2,1} & 1 & r_{2,3} & \cdots & r_{2,p} \\ r_{3,1} & r_{3,2} & 1 & \cdots & r_{3,p} \\ r_{4,1} & r_{4,2} & r_{4,3} & \cdots & r_{4,p} \\ \vdots & \vdots & \vdots & \cdots & \vdots \\ r_{p,1} & r_{p,2} & r_{p,3} & \cdots & 1 \end{bmatrix}$,

$R = BB^t$ where $B$ is the Cholesky factor of $R$ and

$$B = \begin{bmatrix} 1 & 0 & 0 & \cdots & 0 \\ \cos(\theta_{2,1}) & \sin(\theta_{2,1}) & 0 & \cdots & 0 \\ \cos(\theta_{3,1}) & \cos(\theta_{3,2})\sin(\theta_{3,1}) & \sin(\theta_{3,2})\sin(\theta_{3,1}) & \cdots & 0 \\ \cos(\theta_{4,1}) & \cos(\theta_{4,2})\sin(\theta_{4,1}) & \cos(\theta_{4,3})\sin(\theta_{4,2})\sin(\theta_{4,1}) & \cdots & 0 \\ \vdots & \vdots & \vdots & \cdots & \vdots \\ \cos(\theta_{p,1}) & \cos(\theta_{p,2})\sin(\theta_{p,1}) & \cos(\theta_{p,3})\sin(\theta_{p,2})\sin(\theta_{p,1}) & \cdots & \prod_{k=1}^{p-1}\sin(\theta_{p,k}) \end{bmatrix} \text{ for } i > j \text{ angles } \theta_{i,j} \in (0,\pi).$$

To obtain an individual angle $\theta_{i,j}$, we have (where $b_{i,j}$ are the cells of $B$) :

$$\text{For } j = 1 \text{ and } 2 \le i \le p: \;\; \theta_{i,1} = \arccos(b_{i,1}); \text{ and for } 2 \le j < i \le p: \;\; \theta_{i,j} = \text{arcos}\left( b_{i,j} \Big/ \prod_{k=1}^{j-1}\sin(\theta_{i,k}) \right)$$

(17) To obtain an individual correlation, $r_{i,j}$, we have, simply from $R = BB^T$ :

$$r_{i,j} = \cos(\theta_{i,1})\cos(\theta_{j,1}) + \sum_{k=2}^{i-1}\left[\cos(\theta_{i,k})\cos(\theta_{j,k})\prod_{l=1}^{k-1}\sin(\theta_{i,l})\sin(\theta_{j,l})\right] + \cos(\theta_{j,i})\prod_{l=1}^{i-1}\sin(\theta_{i,l})\sin(\theta_{j,l}) \text{ for } 1 \le i < j \le p$$

As stated in Zhang et al. (2015), "Hence a model for the angles ... is equivalent to a model for the correlation matrix." SAS/IML code translating dependence measure values to angle values and angle values to dependence measure values is shown in Table A below. For empirical examples of this translation, see Appendix A, which also includes graphs of the finite sample probability density functions (PDFs) of the angles under various data conditions, along with corresponding spectral distributions.

[20] Note that angle values increase over [0, $\boldsymbol{\pi}$] as dependence measure values decrease over [-1, 1]; or for dependence measures with ranges of [0, 1], the angle values range only over [0, $\boldsymbol{\pi}$/2] and the relationship remains monotonic decreasing.

**TABLE A: Code Translating Correlations to Angles and Angles to Correlations**

**Correlations to Angles**

```
* INPUT rand_R is a valid correlation matrix;

cholfact = T(root(rand_R, "NoError"));

rand_corr_angles = J(nrows,nrows,0);
  do j=1 to nrows;
    do i=j to nrows;
      if i=j then rand_corr_angles[i,j]=.;
      else do;
        cumprod_sin = 1;
        if j=1 then rand_corr_angles[i,j]=arcos(cholfact[i,j]);
        else do;
          do kk=1 to (j-1);
            cumprod_sin = cumprod_sin*sin(rand_corr_angles[i,kk]);
          end;
          rand_corr_angles[i,j]=arcos(cholfact[i,j]/cumprod_sin);
        end;
      end;
    end;
  end;

* OUTPUT rand_corr_angles is the corresponding matrix of angles;
```

**SAS/IML code (v9.4)**

**Angles to Correlations**

```
* INPUT rand_angles is a valid matrix of correlation angles;

Bs=J(nrows, nrows, 0);
do j=1 to nrows;
  do i=j to nrows;
    if j>1 then do;
      if i>j then do;
        sinprod=1;
        do gg=1 to (j-1);
          sinprod = sinprod*sin(rand_angles[i,gg]);
        end;
        Bs[i,j]=cos(rand_angles[i,j])*sinprod;
      end;
      else do;
        sinprod=1;
        do gg=1 to (i-1);
          sinprod = sinprod*sin(rand_angles[i,gg]);
        end;
        Bs[i,j]=sinprod;
      end;
    end;
    else do;
      if i>1 then Bs[i,j]=cos(rand_angles[i,j]);
      else Bs[i,j]=1;
    end;
  end;
end;
rand_R = Bs*T(Bs);

* OUTPUT rand_R is the corresponding correlation matrix;
```

All the above is straightforward and well-established in the literature (see Pourahmadi and Wang (2015)[21]; Ghosh et al. (2021); Rapisarda et al. (2007); Tsay and Pourahmadi (2017); and Zhang et al. (2015)), usually with application to Pearson's correlation matrix. However, the multivariate bijection between the angles and dependence measure matrices holds for all positive definite dependence measures generally. Unfortunately, the relationship between the two multivariate PDFs is nontrivial in the general case, requiring the determinant of the Jacobian matrix, and the finite-sample analytical form of the angles distribution(s) only has been defined for special cases (see Opdyke (2022, 2024a, and 2026) for a foundational example[22]). But this does not prevent us from effectively using the angles space here for inference, because to do so we only need the confidence intervals of the angles under independence, and these can be defined nonparametrically via straightforward and fast simulations (e.g. multivariate uniform data, as described below). This will allow us to utilize an increasingly used and useful parameterization of positive definite dependence measures, especially in financial settings, with a long list of well-established advantages (i.e. the angles space i. is generally more robust (see Lucchetti and Pedini (2023)), especially under heavy-tailed data (see Zhang and Yang (2025)); ii. allows for unconstrained optimization (see Zhang et al. (2015)) that even can be freely characterized via regressions as functions of covariates; iii. provides better estimation of such structures (see Pinheiro and Bates (1996) and Lucchetti and Pedini (2023)). So we use the angles space here, which is, by design and definition, a positive definite space (the Cholesky factor is undefined otherwise), for simultaneous ***inference*** of all the pairwise relationships. And when these inferences are based on directional dependence measures within the framework of dual orderings, they will allow ADD to establish the ***direction*** of any such relationships.

---

[21] Note a typo in Pourahmadi and Wang's (2015) unnumbered equation (close to (2)) that is the analogue to (16) above, where they insert $\prod_{k=1}^{j-1} \sin\left(\arccos\left(b_{i,k}\right)\right)$ instead of $\prod_{k=1}^{j-1} \sin\left(\theta_{i,k}\right)$. This same typo was noted in the errata of Lucchetti and Pedini (2025).

[22] Using a spherical angles parameterization, Opdyke (2022, 2024a, and 2026) provides the complete finite sample analytic distribution, including both the matrix-level and cell-level PDFs, CDFs, and inverse CDFs, of Pearson's matrix under the Gaussian identity matrix (see www.DataMineit.com/DMI_publications.htm for an interactive, downloadable excel workbook

## 3. Methodology

### 3.A. Rationale: Dual Orderings Using the Angles Space of Directional Dependence Measures

We now rely on the above descriptions of directional dependence measures and the (positive definite) angles space to outline the steps implementing ADD. We note, again, that ADD is designed specifically for causal discovery: it is not a method designed to provide effect sizes, as do regression-based approaches (see for example Pascual-Marqui et al. (2024); Blömbaum et al. (2019); and MacKinnon & Lamp (2021)), nor does it provide counterfactual outcomes. Rather, ADD identifies existing causal structure aimed at DAG recovery.

The approach here is to use, like some of the other causal frameworks listed above, the directional dependence measures described above in section 2. Take, for example, Chatterjee's improved correlation ("ICH") coefficient (see Xia et al. (2025)), which ADD can use twice via dual orderings, once with the data variables in one column order, and again with the data variables in the reverse column order. So if we have a treatment variable (X) and a dependent variable (Y) and relevant covariates (V1, V2, V3), we column-sort the data in one order (e.g. an ordering of X, V1, V2, V3, Y), and then column-sort the data in the reverse order (Y, V3, V2, V1, X), and apply ICH to each ordering; the two resulting ("forced-symmetric") ICH matrices will together capture all potential associations, in both directions, of all the variables/factors.[23] And all the cells of these two estimated dependence matrices will fully map to the relevant variable categories that make up a DAG (e.g. the confounders, colliders, mediators, independent variables, causes of X, consequences of X, causes of Y, and consequences of Y). For example, for a implementing this). Joarder and Ali (1992) were the first to provide the matrix level PDF under these same conditions. Importantly, note that the angles are independent random variables under these specific, and some broader, conditions.

[23] The same rationale applies to feature selection generally, which in quantitative finance often is referred to as feature engineering.

particular pair of factors, X and Y (i.e. the X-Y cell of the matrix, which is equivalent to an edge in a graph), two findings from this approach – one of a statistically significant, directional effect of X on Y (X→Y), but NO statistically significant, directional effect of Y on X (Y→X) – provide evidence of a directional effect of X on Y (X→Y). When assumptions A1-A7 are satisfied (see Appendix C), under conditioning such evidence is consistent with adjacency, and under marginal dependence it is consistent with reachability. For readability purposes, we leave the formal definitions of A1-A7 in Appendix C,[24] and present the rationale for this approach here in the text as follows.

Consider a parent–child pair X→Y with $Y = f(X, U_Y),\ U_Y \perp X$, and NOT Y→X. Under additive-noise mechanisms, $Y = f(X) + U_Y$, independence between X and $U_Y$ implies that the conditional variability of Y given X is dominated by $U_Y$, whereas the inverse mapping $X = g(Y) + \varepsilon$ generally violates independence (i.e. Y and the induced noise are statistically dependent). Independence between the cause and its exogenous noise term is the identifying restriction that generates directional asymmetry: only the true causal direction preserves independence, while the inverse mapping generally will violate it. This violation is precisely what ADD's directional measures exploit when evaluating the two possible directions of a pairwise relationship. In other words, because the one direction preserves independence between the input and the associated exogenous noise, while the reverse generally does not, the dual-order calculations allow the expected asymmetry to manifest directly in the empirical directional measures. So we therefore have that directional functionals *M* that reward predictability under independent noise yield $E\left[M(X \rightarrow Y)\right] > E\left[M(Y \rightarrow X)\right]$.

Now for linear-Gaussian cases, this asymmetry collapses as both directions are equivalent. But for nonlinear and/or non-Gaussian and/or heteroskedastic mechanisms, the asymmetry is restored. And ADD leverages this by: (i) computing *M* in both directions (via dual orderings), (ii) performing joint finite-sample statistical tests in angle space under positive definite constraints, and (iii) calling an edge only when one direction is significant and the other direction is not. We implement these steps, using ICH, in the preliminary empirical study described below.

## 4. Preliminary Empirical Study

The objective of ADD and this study is to identify causal structure consistent with the 'ground truth' DAG, assuming it is rightly specified.[25] We demonstrate ADD's effectiveness for this purpose in the empirical study presented below, and importantly note again that, as explained in detail in Appendix C, under assumptions A1-A7, and when directional dependence is evaluated conditionally on DAG space V \ {X, Y}, ADD's one-direction-only rule is pointwise sound for adjacency; when directional dependence is evaluated marginally, ADD identifies directed reachability rather than direct parenthood. The empirical study presented herein applies conditioning, thus allowing for interpretation under the former (direct adjacency). Promising results from this preliminary study will justify more expansive testing in a larger, follow-up study described below. For the reader's reference during this step-by-step explanation, we place ADD's entire DAG recovery pipeline here in Graph 1.

[24] Again, the major assumptions include acyclicity, causal sufficiency, Markov and faithfulness, independent and identically distributed sampling, reliable measurement, and monotone link from (directional) dependence measure to asymmetry.

[25] MacKinnon and Lamp (2021) appropriately advise that "The correct causal model is an exacting qualification, requiring a program of research with precise definition of causal effects, specification of assumptions, and sensitivity analysis for how violating assumptions affects results. Statistical analysis is useful for demonstrating associations between variables that are consistent or inconsistent with a causal model."

We start with the straightforward, but non-toy DAG presented in Digitale et al. (2022) (see Graph 2 below), which contains confounders, colliders, and mediators, as well as both causes and consequences of the two primary factors of interest, E and D. There are multiple causal paths linking E and D (E→D, E→M→D) as well as multiple non-causal paths linking E and D (E←C→G→D, E→S←L→D). Rightly identifying the directional relationships shown via the arrows in the DAG – and ONLY these relationships – will allow researchers to obtain unbiased estimates of the effect of E on D (by controlling for other effects on D, but only where appropriate). So accurate DAG recovery is the necessary precursor to this estimation. Note that some of the roles these variables play include C as a 'confounder,' M as a 'mediator,' S as a 'collider,' Z as an effect of D, I as an instrumental variable, i.e. a cause of E but not (directly) of D, J as an effect modifier, etc. For a complete description of the DAG we refer readers to Digitale et al. (2022).

**Graph 1: DAG Recovery Pipeline**

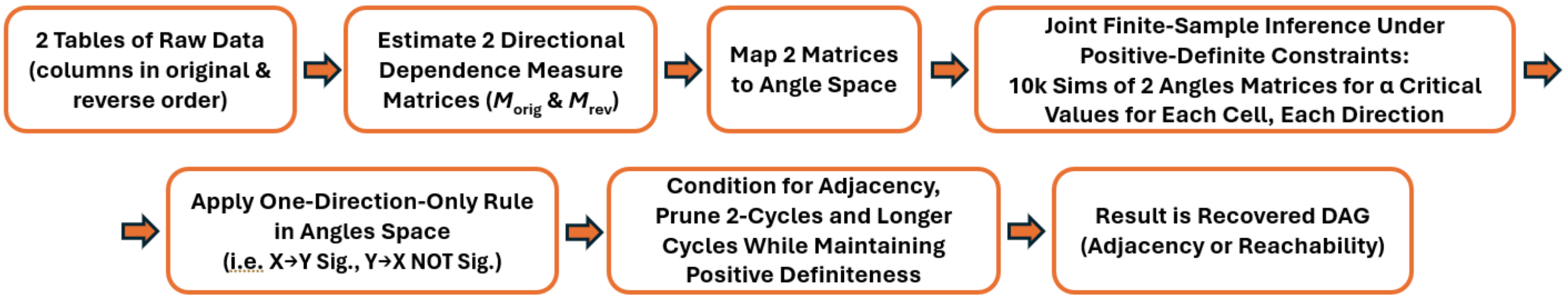

**Graph 2: Directed Acyclic Graph of Digitale et al. (2022)**

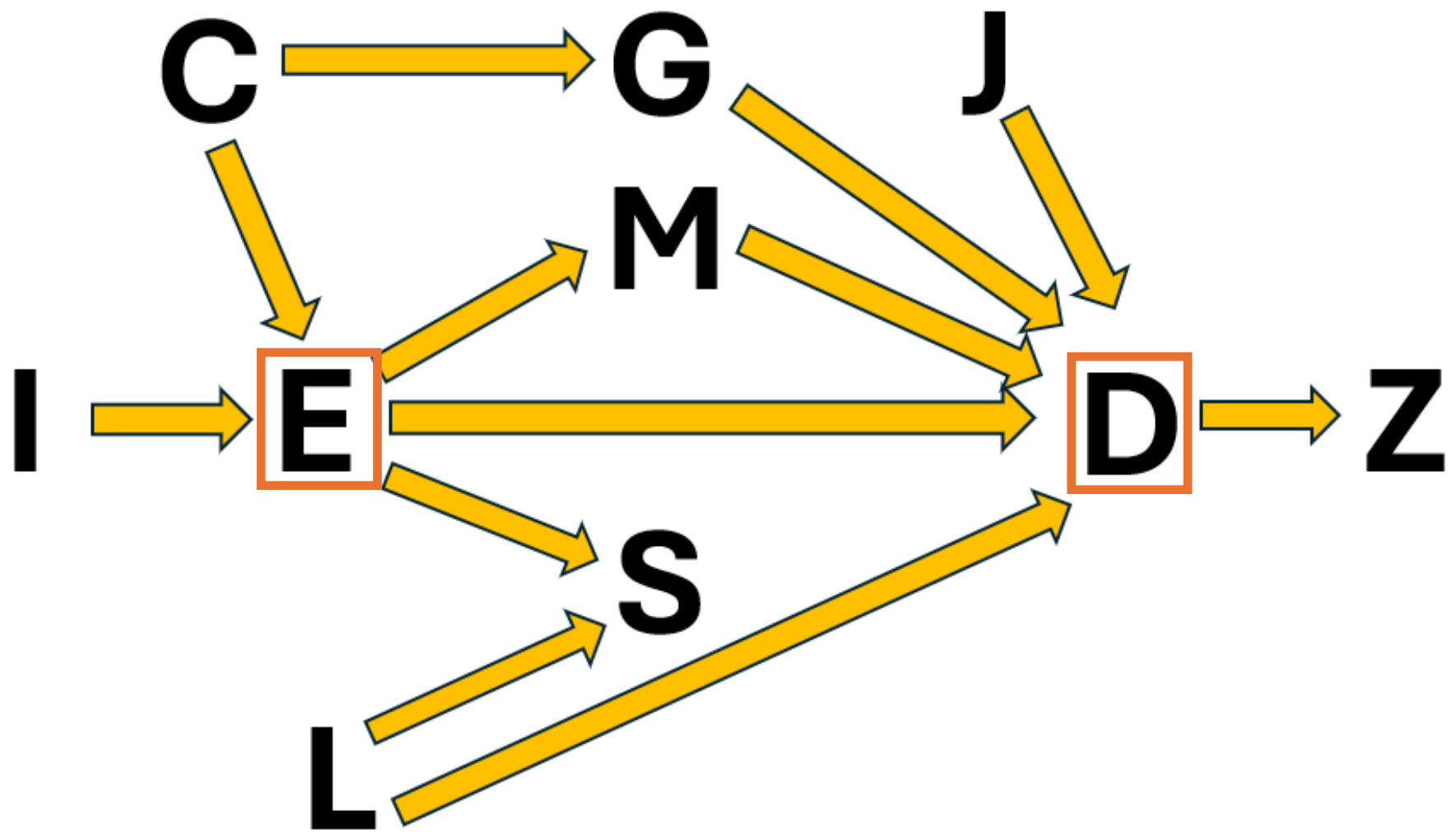

The dimension of the corresponding all-pairwise matrix, where each cell represents an edge of the graph, is p = 10 factors, making p(p-1)/2 = 45 pairwise cells. ADD's objective here is to rightly identify the directional effect in each of the 12 cells where one exists, and to rightly identify no effect in the 33 cells that have no effect (see Graph 3). We generate data under the null hypothesis of independence, to obtain the relevant (one-sided) confidence intervals, using a multivariate uniform distribution with independent components (only independence matters here; distributional changes make virtually no empirical difference in the distributions of the angles under the null, as expected).[26] We generate data based on the alternate hypothesis (the DAG) via the code in Table B below.[27] All parametric distributions are heavy-tailed (Student's t-distributed),[28] all error terms are asymmetric (lognormally distributed), and all functional relationships are highly nonlinear (most are sinusoidal, with some polynomial terms). As under the null, every simulation has n = 100 observations (which represents only about five months of trading days).[29] This yields, over 10,000 simulations of this DAG, the average Pearson's, Spearman's, and

---

[26] Under independence, the sampling distributions of dependence measures, and hence, those of the angles based on them, are not functions of the underlying marginal data distributions. The reasons for this are similar and overlapping across dependence measures. For Pearson's rho, the sampling distributions are functions of sample size (n) and kurtosis, but the latter vanishes for non-diagonal cells under independence, making Pearson's density strictly a function of n. For Kendall's tau and Spearman's rho, the margin-free property holds under dependence as well because their use of rankings discards all information regarding the marginal densities. Specifically under independence, the ranks all are uniformly distributed. In either case, independence or dependence, no marginal information is used by these measures when computing pairwise dependence, and the sampling distributions again are strictly a function of n. Finally, using two directional matrices based on Chatterjee's or improved Chatterjee's (or similar methods like DDC) as described herein, the same argument holds: the rankings these measures use discard any marginal information in the pairwise computation, and the distributions of the two ordered cell values ( $\xi_n(X,Y)$ and $\xi_n(Y,X)$ ) are identical and based only on n, as is their joint distribution. Geometric arguments also confirm the margin-free property of these pairwise dependence measures.

Therefore, the sampling distributions of the angles, which are based on these dependence measures, do not depend in any way on the marginal densities of the underlying data: only independence matters when simulating the multivariate data distribution to calculate the confidence intervals of the angles distributions (under independence). So to scale with maximal efficiency in the empirical study herein, we simulate an independent multivariate uniform distribution to obtain the confidence intervals for the angles.

[27] One of the goals here was to test the power to identify the individual effects of multiple parent variables, with different frequencies and functional forms, on a single child variable. The extensive use of coefficients in Table B allowed for such testing (e.g. the individual effects of J, E, M, G, and L on D) by preventing differences in scale from compounding by orders of magnitude. Performance under standardized scale is consistent with an absence of the marginal variance problem identified by Reisach et al. (2021). Unlike other causal discovery algorithms, ADD does not appear to have this problem, as we might expect due at least in part to the Cholesky factor placing angles on the UNIT hyper-hemisphere. However, this will be explicitly tested in a follow-up empirical study.

[28] Notably, angles-based dependence measures have been shown to be very robust to heavy-tailed data (see Zhang and Yang (2025)).

[29] Using reasonably small samples is conservative here, making it more difficult for ADD to rightly identify directional effects. Strong results under these conditions indicate good power of the estimator. Note that other studies treating DAG recovery use sample sizes as large as n=1,000 (see Olivetti et al. (2026) and https://hub.crunchdao.com/competitions/causality-discovery), and even n=4100 (see Xue et al. (2025)) and n=5000 (see Petkov et al. (2025)).

**Graph 3: Directional Effects of 12 Pairwise Relationships**
**(Effect Direction is Column to Row, Original Order)**

ORIGINAL ORDER

| | I | C | E | G | L | S | M | J | D | Z |
|---|---|---|---|---|---|---|---|---|---|---|
| I | | | | | | | | | | |
| C | X | | | | | | | | | |
| E | ← | ← | | | | | | | | |
| G | X | ← | X | | | | | | | |
| L | X | X | X | X | | | | | | |
| S | X | X | ← | X | ← | | | | | |
| M | X | X | ← | X | X | X | | | | |
| J | X | X | X | X | X | X | X | | | |
| D | X | X | ← | ← | ← | X | ← | ← | | |
| Z | X | X | X | X | X | X | X | X | ← | |

REVERSE ORDER

| | Z | D | J | M | S | L | G | E | C | I |
|---|---|---|---|---|---|---|---|---|---|---|
| Z | | | | | | | | | | |
| D | → | | | | | | | | | |
| J | X | → | | | | | | | | |
| M | X | → | X | | | | | | | |
| S | X | X | X | X | | | | | | |
| L | X | → | X | X | → | | | | | |
| G | X | → | X | X | X | X | | | | |
| E | X | → | X | → | → | X | X | | | |
| C | X | X | X | X | X | X | → | → | | |
| I | X | X | X | X | X | X | X | → | X | |

Kendall's matrices shown in Table C. These show essentially no non-directional associations.[30] When implementing ADD here we chose the improved Chatterjee correlation coefficient ("ICH") of

[30] One of the cells shows a slight effect, but only under Pearson's (which is more sensitive to outliers), and this is almost certainly due to the different functional form for that factor, which is one (a quadratic term) that is more sensitive to bi-directional effects (when using ICH) than those characterizing most of the DAG (i.e. sinusoidal).

**TABLE B: SAS/IML Code Generating Data Under the Alternate Hypothesis (DAG)**

```
call randgen(R_ERR_LOGN, "LOGNORMAL",0,sqrt(log( (1+sqrt(5))/2 )) );
R_ERR_LOGN = R_ERR_LOGN - exp( log( (1+sqrt(5))/2 )/2 );

call randgen(R_ERR_LOGN2, "LOGNORMAL",0,sqrt(log( (1+sqrt(5))/2 )) );
R_ERR_LOGN2 = R_ERR_LOGN2 - exp( log( (1+sqrt(5))/2 )/2 );

call randgen(R_ERR_LOGN3, "LOGNORMAL",0,sqrt(log( (1+sqrt(5))/2 )) );
R_ERR_LOGN3 = R_ERR_LOGN3 - exp( log( (1+sqrt(5))/2 )/2 );

call randgen(R_ERR_LOGN4, "LOGNORMAL",0,sqrt(log( (1+sqrt(5))/2 )) );
R_ERR_LOGN4 = R_ERR_LOGN4 - exp( log( (1+sqrt(5))/2 )/2 );

call randgen(R_ERR_LOGN5, "LOGNORMAL",0,sqrt(log( (1+sqrt(5))/2 )) );
R_ERR_LOGN5 = R_ERR_LOGN5 - exp( log( (1+sqrt(5))/2 )/2 );

call randgen(R_ERR_LOGN6, "LOGNORMAL",0,sqrt(log( (1+sqrt(5))/2 )) );
R_ERR_LOGN6 = R_ERR_LOGN6 - exp( log( (1+sqrt(5))/2 )/2 );

call randgen(CCC, "T", 10);

call randgen(III, "T",10);

EEE = cos(2*constant('PI')*CCC) + cos(2*constant('PI')*III) + R_ERR_LOGN/8;

GGG = -sin(2*constant('PI')*0.95*CCC) + R_ERR_LOGN2;

call randgen(LLL, "T",10);

SSS = -cos(2*constant('PI')*LLL) - sin(2*constant('PI')*(EEE)) + R_ERR_LOGN3;

MMM = -0.7*cos(2*constant('PI')*(0.95*EEE)) + R_ERR_LOGN4/3;
call randgen(JJJ, "T",10);

DDD = -2.3*cos(2*constant('PI')*(0.6*JJJ)) + 1.35*cos(2*constant('PI')*(0.25*EEE)) - 3.8*MMM##2 +
2.1*cos(2*constant('PI')*(0.65*GGG)) + 2.3*sin(2*constant('PI')*LLL) + R_ERR_LOGN5/8;

ZZZ = cos(2*constant('PI')*(DDD)) + R_ERR_LOGN6/6;
```

Xia et al. (2025),[31] and because this is a directional measure whose values range from zero (independence) to one (perfect dependence), we need specifically the upper (1 – α) confidence bound[32] under data generated under independence. This is needed to test whether one can reject the null hypothesis of independence when the data are generated under the alternate hypothesis (the DAG).[33] We set $\alpha = 0.05$, but conservatively adjust it using a Bonferroni adjustment ($\alpha_adj = \alpha/2$) because we are conducting two tests on the same data for each edge / cell of the matrix: one with the columns of factor data in the "original" order, and one with the columns of factor data in the reverse order (see Graph 3; the order of the columns in the data matches the column order(s) in the matrices).[34]

---

[31] ICH is noted in Xia et al. (2025) to have good power under these data conditions, so it is an appropriate choice for a preliminary study whose design is meant to test and justify further inquiry (we propose below an extensive follow-up study under wide-ranging data conditions). But this study poses nontrivial challenges to testing the efficacy of ADD in other ways as well, such as using much smaller sample sizes compared to other studies cited herein. The choice of dependence measure also brings up a potentially large advantage of ADD over competitors, that is, its ability to opportunistically use any (positive definite) directional dependence measure as befits the data conditions of different settings. This flexibility can provide increases in statistical power that other methods based on fixed measures will lack (especially if ADD is used with 'stacking,' as discussed below). In fact, the data conditions of this study are more consistent with a setting we want to focus on – quantitative finance – than others tested in the literature (e.g. circular data), so it is reasonable, and a benefit of ADD, to be able to select an appropriate dependence measure that acknowledges and exploits the type of data we are faced with in a given setting.

[32] The upper confidence bound is (1 – α) rather than (1 – α/2) because this is a one-sided hypothesis test.

[33] Recall that ADD's inference is based on the angles corresponding to each cell of the matrix, so this is actually the lower α bound because of the monotonic decreasing relationship between angles and dependence measure values: angles approach zero as dependence measure values approach one (or perfect dependence), and angles approach π/2 as (directional) dependence measure values approach zero (or independence). For non-directional measures, like Pearson's or Kendall's or Spearman's, angles approach π as dependence measure values approach negative one, or rather, the lower bound of the matrix, as negative one is a maximal lower bound that is not always attainable.

[34] Under the null hypothesis of independence these two tests will be independent, and under the alternate hypothesis these two tests will be, on average, negatively correlated. The Bonferroni adjustment (see Bonferroni (1936)) is used here not only because it is very conservative but also because it controls Familywise Error Rate (FWER) regardless of the direction of any correlation between these two tests. The effects of multiplicity across cells are ignored for purposes of this exercise.

**TABLE C: Average Pearson's Rho, Spearman's Rho, and Kendall's Tau Matrices**
**10,000 Simulations Under the Alternate Hypothesis of the DAG**

**Pearson's Rho**

| | I | C | E | G | L | S | M | J | D | Z |
|---|---|---|---|---|---|---|---|---|---|---|
| I | 1 | | | | | | | | | |
| C | 0.001 | 1 | | | | | | | | |
| E | -0.001 | 0.000 | 1 | | | | | | | |
| G | -0.002 | -0.002 | -0.001 | 1 | | | | | | |
| L | 0.000 | -0.002 | -0.001 | 0.001 | 1 | | | | | |
| S | -0.001 | 0.001 | 0.052 | 0.001 | -0.003 | 1 | | | | |
| M | 0.000 | 0.000 | 0.006 | 0.001 | 0.000 | 0.000 | 1 | | | |
| J | 0.000 | -0.001 | 0.000 | -0.001 | -0.003 | 0.000 | 0.000 | 1 | | |
| D | 0.001 | 0.001 | 0.000 | -0.042 | 0.000 | 0.001 | -0.166 | 0.000 | 1 | |
| Z | -0.001 | 0.001 | 0.000 | 0.000 | 0.001 | 0.000 | -0.002 | -0.002 | -0.001 | 1 |

**Spearman's Rho**

| | I | C | E | G | L | S | M | J | D | Z |
|---|---|---|---|---|---|---|---|---|---|---|
| I | 1 | | | | | | | | | |
| C | 0.001 | 1 | | | | | | | | |
| E | -0.001 | 0.000 | 1 | | | | | | | |
| G | -0.002 | -0.002 | -0.001 | 1 | | | | | | |
| L | 0.000 | -0.001 | -0.001 | 0.001 | 1 | | | | | |
| S | 0.000 | 0.000 | 0.039 | 0.001 | -0.002 | 1 | | | | |
| M | 0.000 | 0.000 | 0.004 | 0.001 | 0.000 | -0.001 | 1 | | | |
| J | 0.000 | -0.002 | 0.000 | 0.000 | -0.004 | 0.001 | 0.001 | 1 | | |
| D | 0.000 | 0.000 | 0.000 | -0.054 | -0.001 | 0.001 | -0.038 | -0.001 | 1 | |
| Z | -0.001 | 0.001 | 0.000 | 0.000 | 0.001 | -0.001 | -0.002 | -0.001 | -0.001 | 1 |

**Kendall's Tau**

| | I | C | E | G | L | S | M | J | D | Z |
|---|---|---|---|---|---|---|---|---|---|---|
| I | 1 | | | | | | | | | |
| C | 0.001 | 1 | | | | | | | | |
| E | -0.001 | 0.000 | 1 | | | | | | | |
| G | -0.001 | -0.001 | -0.001 | 1 | | | | | | |
| L | 0.000 | -0.001 | -0.001 | 0.001 | 1 | | | | | |
| S | 0.000 | 0.000 | 0.026 | 0.001 | -0.002 | 1 | | | | |
| M | 0.000 | 0.000 | 0.003 | 0.000 | 0.000 | 0.000 | 1 | | | |
| J | 0.000 | -0.001 | 0.000 | 0.000 | -0.002 | 0.001 | 0.000 | 1 | | |
| D | 0.000 | 0.000 | 0.000 | -0.037 | -0.001 | 0.001 | -0.025 | 0.000 | 1 | |
| Z | -0.001 | 0.001 | 0.000 | 0.000 | 0.001 | -0.001 | -0.001 | -0.001 | -0.001 | 1 |

For each cell, there are four possible outcomes of the two tests:
A. significant effect in the original direction, no significant effect in the reverse direction
B. no significant effect in the original direction, significant effect in the reverse direction
C. no significant effect in either direction
D. significant effects in both directions

Under independence, the probabilities associated with A.-D. then become:
A. Pr(A.) = $\alpha$_adj * (1 – $\alpha$_adj) = 0.025*0.975 = 0.024375
B. Pr(B.) = $\alpha$_adj * (1 – $\alpha$_adj) = 0.025*0.975 = 0.024375
C. Pr(C.) = (1 – $\alpha$_adj) * (1 – $\alpha$_adj) = 0.975*0.975 = 0.950625
D. Pr(D.) = ($\alpha$_adj) * ($\alpha$_adj) = 0.025*0.025 = 0.000625

After estimating two dependence measure (ICH) matrices, one for each column-ordering ('original' and 'reverse'), we apply conditioning as described in Appendix D. Next, any 2-cycles from category D. above are pruned to preserve the acyclicity of the DAG by retaining only the stronger of the two effects (i.e. the one with the smaller p-value). Finally, longer cycles are pruned based on several criteria including the number of called edges (cells) per node (factor/variable), the sum of the p-values associated with all the nodes with the same number of edges, and within each of these groups, the size of the individual p-values of the specific edges (see Appendix C for details). Importantly, note that all conditioning and pruning are performed under the constraint of positive definiteness. Now, with conditioning applied and acyclicity enforced, what we are interested in is the random variable that is the number of the 45 pairwise cells, which are edges of the DAG, that are rightly identified, where rightly identified means the 12 cells with directional relationships are rightly chosen as outcome A., and those 33 cells with no directional relationships are rightly chosen as outcome C. Under the null hypothesis of independence, this distribution is shown in Graph 4 below, and the expected value is approximately 31.66 right identifications.[35]

[35] We obtain this distribution via a simple simulation that applies random uniform variables generated with support [0, 1] to rightly identify each of the 12 directional relationships with Pr(A.) and each of the 33 non-relationships with Pr(C.) yielding, after 100 million simulations, Graph 4, with an expected value of approximately 31.66 right identifications.

**Graph 4: Distribution of #Rightly Identified Directional Effects/Non-Effects: All 45 Pairs (Fully Exploratory Analysis, No Priors)**

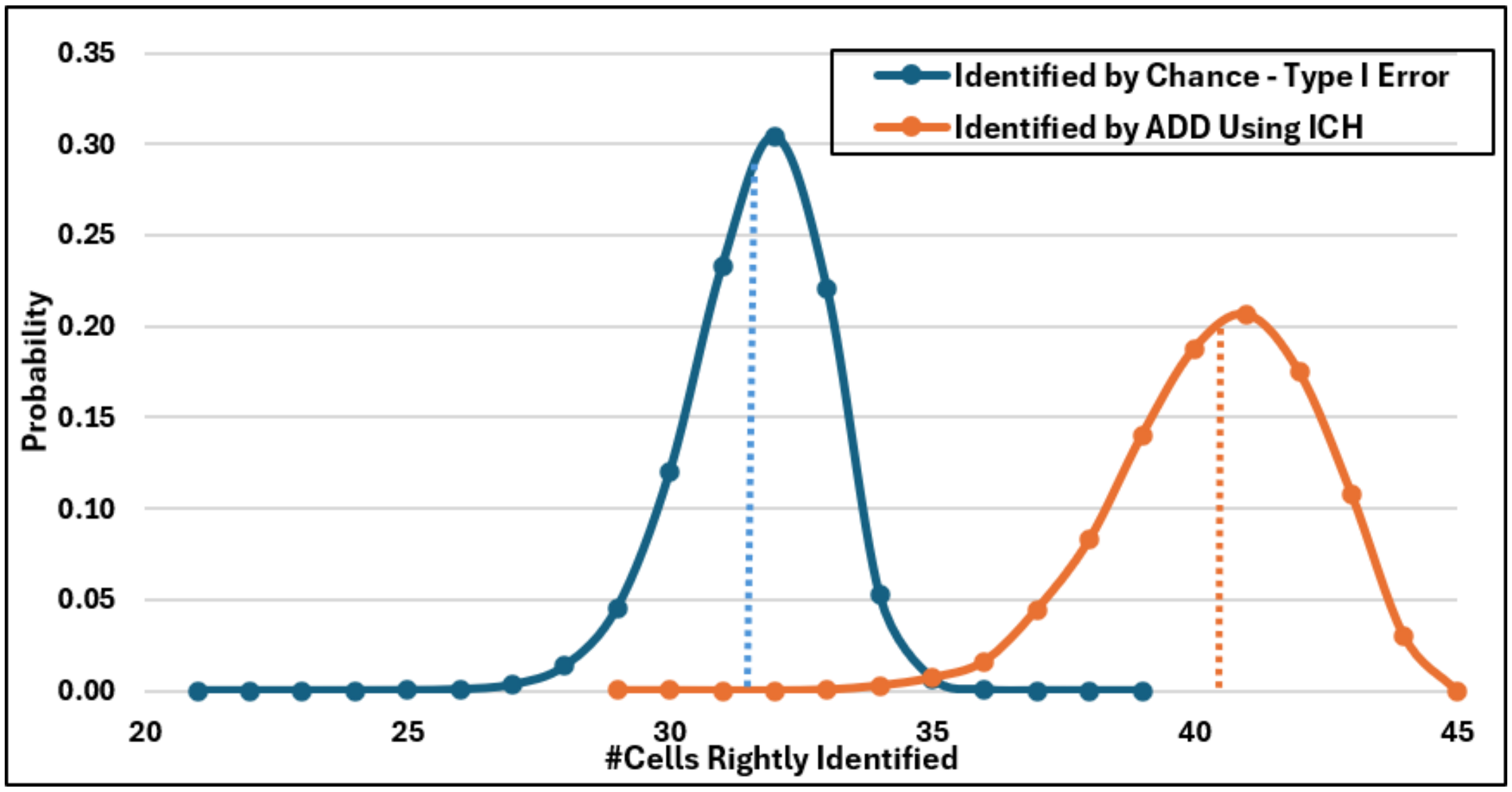

ADD is applied as described above and shown in the two matrices in Graph 3. The columns/factors of data first are sorted in order so that the 12 relationship cells all are in the 'right' direction according to the DAG. Graph 3 in the Original Order shows the directional effect going from column to row (so, for example, the cell with row E and column I indicates the effect of factor I on factor E). ICH is applied to the data with the factors in this order, and the resulting matrix of ICH values is saved. Then the column order is reversed, ICH is again applied, the resulting matrix of ICH values is saved, and then both the rows and columns of this second matrix are reversed so that the two matrices can be compared easily, cell for cell. If, for a given cell, the result of the first test is statistically significant, but the result of the second test is NOT statistically significant, then the result is classified as outcome A.: a directional effect in the right direction. For the 33 'non-relationship' cells, a 'right' outcome is a finding of no statistically significant effect in either direction, i.e. outcome C. This is repeated 10,000 times, and for each simulation, conditioning is applied, acyclicity is enforced (both while preserving positive definiteness), and then the number of times each cell is rightly classified, either as one with a directional effect or one with no effect, is summed and shown in Graph 4 above. The expected value of this distribution is 40.44, compared to that under the null which is 31.66. The difference between the two distributions, which unarguably is material, also is highly statistically significant, with a p-value<0.00001.[36] Note that this #right metric simply is the complement (relative to total # edges=45) of the Structural Hamming Distance (SHD), that is, the number of edge modifications needed for an estimated DAG to match the true DAG. The distribution of the corresponding SHDs across the 10,000 simulations, with a mean of 4.56, is shown in Graph 4a. We combine Graphs 4 and 4a in Graph 4b to put them on the same scale.

[36] This is a one-sided p-value based on a permutation test of 100,000 samples with sample sizes of n1=n2=45 (see Opdyke (2003, 2011, and 2013) for details on permutation tests). These sample sizes are used because they reflect the outcomes of 45 Bernoulli variables, that is, for 45 cells/edges, 45 outcomes of zero or one that sum to the #right in a given sample.

In addition to SHD, we present the confusion matrix, otherwise known as the error matrix, below in Table D, along with various common model performance metrics based on it.

**Graph 4a: Distribution of SHD: #Edges Changed to Match True DAG**

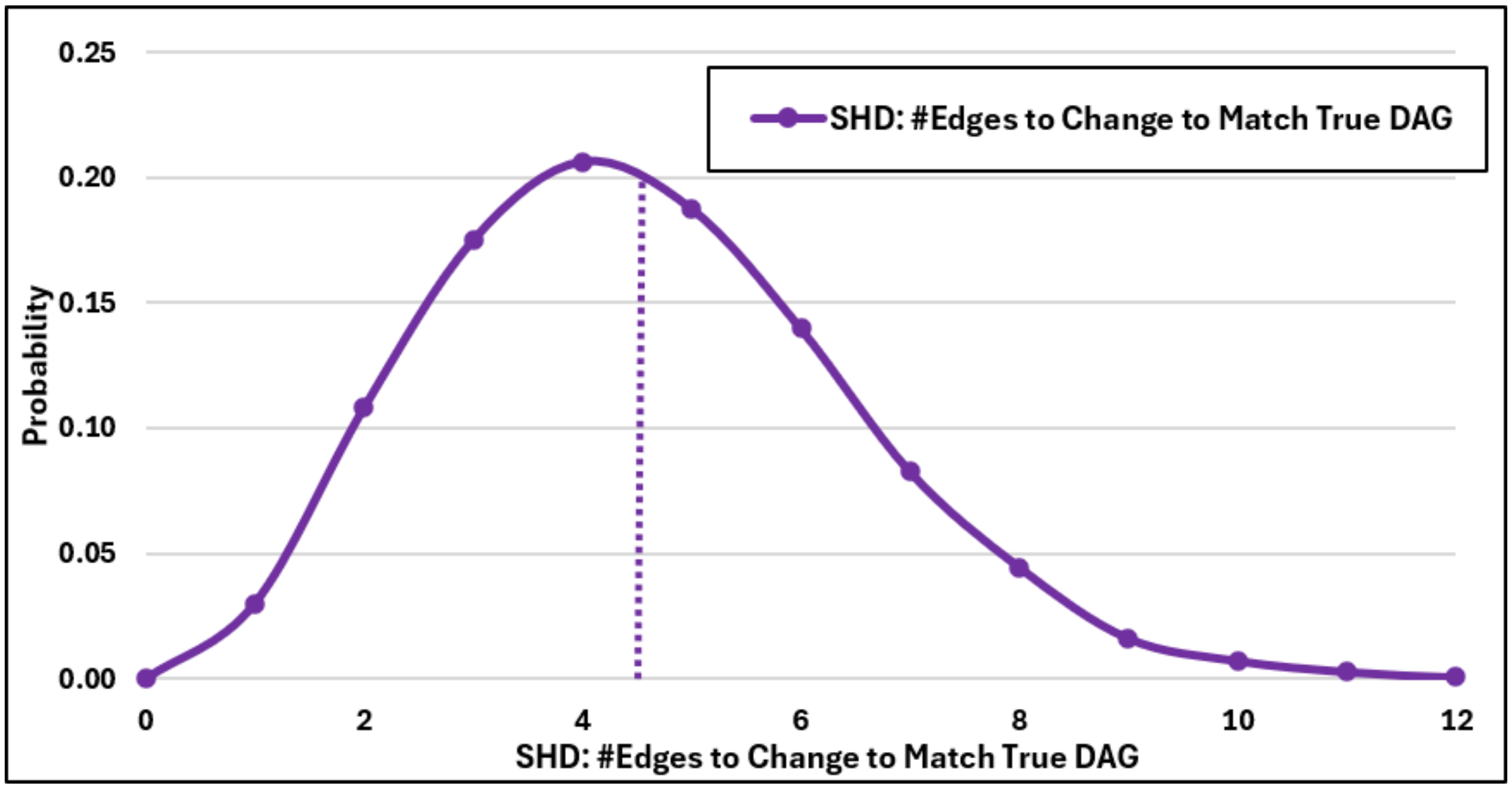

**Graph 4b: Distribution of #Rightly Identified Directional Cells/Edges vs. SHD**

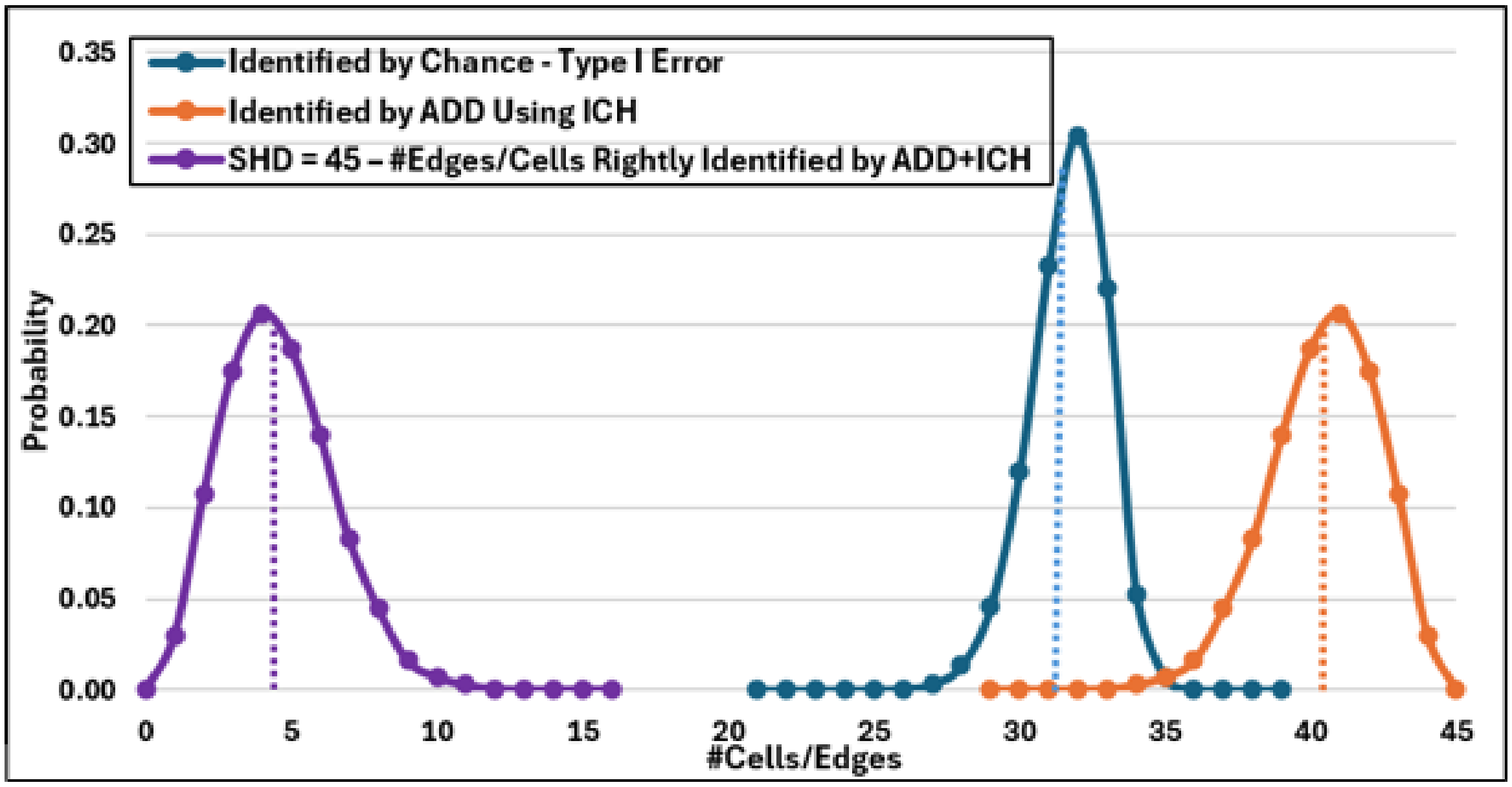

**TABLE D: Error Matrix, and Associated Performance Metrics**

**Error Matrix - Numbers**

| | Predicted Right (Edge+Dir) | Predicted Not Right (Edge+Dir) | TOTAL |
|---|---|---|---|
| True Edge | 80,949 | 39,051 | 120,000 |
| Non-Edge | 6,552 | 323,448 | 330,000 |

**Error Matrix - Rates**

| | Predicted Right (Edge+Dir) | Predicted Not Right (Edge+Dir) | TOTAL |
|---|---|---|---|
| True Edge | 67.5% | 32.5% | 100% |
| Non-Edge | 2.0% | 98.0% | 100% |

| | | | | | |
|---|---|---|---|---|---|
| Precision = | 0.925 | MCC = | 0.732 | Jaccard = | 0.640 |
| Recall = | 0.675 | F1 = | 0.780 | Prevalence Threshold= | 0.146 |
| Accuracy = | 0.899 | Youden = | 0.655 | Balanced Accuracy = | 0.827 |

To briefly review the aggregate results from Table D, Recall is 67.5%, which is reasonably strong given challenging data and relatively small sample sizes.[37] This is accompanied by a stronger F1 score at 78%, and an even stronger Accuracy at 89.9%. We avoid the debate in the literature regarding balanced vs. unbalanced aggregate performance metrics and simply present that ADD shows a strong Balanced Accuracy of 82.7%, while Precision is even stronger at 92.5%. The metric we place the most weight on herein, however, is Matthews Correlation Coefficient (MCC) because it not only uses all the information in the error matrix (as opposed to just some of the aggregated cell values), but also has a number of methodological advantages over most other measures.[38] By this measure (73.2%) the model discriminates quite well in rightly identifying true, directly adjacent, directional effects in the DAG. Note that all of these metric values are consistent with our conservative application of ADD: because the same data is tested twice, once in each direction, the (conservative) Bonferroni adjustment decreases both the number of true positives (thus lowering Recall) and the number of false positives (thus increasing Precision). While the type 1 vs. type 2 error tradeoff can never be avoided altogether with sample data,[39] quantitative tradeoffs between the two remain functions of implementation and judgement based on the specific setting of application. For ADD's implementation herein, we chose conservatism when addressing the multiplicity inherent in statistical tests under dual orderings; yet the aggregate empirical results of this preliminary study remain promising even under challenging data conditions and non-large sample sizes, easily justifying and motivating a more extensive follow-up study.

[37] A recent online DAG recovery contest, with over 1,900 registered users and over 3,300 submissions (see Olivetti et al. (2026) and https://crunchdao.com/case-studies/adia-lab-causality), used sample sizes (n=1,000) ten times as large as those used herein (see https://hub.crunchdao.com/competitions/causality-discovery). Sample sizes in other studies (see Xue et al. (2025) and Petkov et al. (2025)) are larger still (n=4,100 and n=5,000, respectively).

[38] Itaya et al. (2024) cite MCC as "a performance metric with high reliability, offering a balanced measurement even in the presence of class imbalances" while Stoica and Babu (2023) describe it as "one of the most useful metrics for the performance of a binary classification or hypothesis testing method … [providing] a more informative and thus more useful performance score than many of the other metrics."

We also present below in Table E the percentage of right classifications, across all simulations, for each specific cell/edge. Cells with red font indicate lower relative scores on the %right metric. The red on the effect of M on D is almost certainly due to the functional form being a quadratic term as opposed to a sinusoidal term, where the former is more sensitive to bi-directional results (when using ICH) and thus, yields fewer rightly identified one-direction effects; also, the effect of E on D is muted due to the large number of parent nodes to D, which makes this edge more challenging to rightly identify. This type of table can be very informative regarding the specific strengths and weaknesses of an estimator and its framework under different DAGs and different data conditions.

If we examine only the 12 cells with directional relationships, the results remain strong. Under the null hypothesis of independence, the distribution of the number of right identifications becomes simply a binomial distribution with probability = Pr(A.), because there are no cells with 'no relationship' included. This is shown in Graph 5, compared to ADD's distribution of the number of right identifications for these 12 cells. The expected value of the former is approximately 0.29, and of the latter, 8.09. This very material difference also is highly statistically significant, even under smaller sample sizes (n1=n2=12), with a p- value < 0.00001.[40]

Even if we gave a researcher an unfair advantage over ADD and he or she somehow KNEW that the 12 cells were associated with an effect (i.e. perfect priors), without knowing the direction of the effect, the distribution under the null would be a coin toss, with an expected value for the number of right selections of 6 out of the 12. Even under these conditions (see Graph 6), ADD's effect is material, and statistically significant, with a p-value of just over 0.01. All of these results are summarized below in Table F.[41]

[39] Of course, the aim of developing more powerful statistics is to mitigate, as much as possible, both simultaneously.

[40] For this permutation test, sample sizes n1 = n2 = 12, and 100,000 simulations were run.

[41] It is worth noting here that, as a mathematical issue, the values of the Cholesky factor do depend on the order of the columns of the underlying data. However, in this setting (i.e. all-pairwise dependence measure matrices that are unit diagonal, positive definite, and not extremely sparse or on scales that differ by orders of magnitude) this appears to be more of a computational/calculation issue than a statistical one.

The conditions under which the Cholesky factor's values will change materially based on changing column order include: the pathological case of the dependence measure matrix being very nearly singular (e.g. nearly perfectly collinear variables), the matrix being extremely sparse (i.e. the vast majority of cells are zeros), and/or the matrix contains values that are highly non-uniform (e.g. with large clusters of values at or very near zero, and large clusters of values at or very near one). These data conditions are relatively extreme and readily observed, so if the degree to which they exist is a concern, the potential materiality of the effects of column ordering can be checked empirically. We perform such a check based on the empirical study performed in the paper, randomly re-ordering the columns of the input data variables ten different ways (and changing nothing else): we find that, compared to the original findings from Table D, the average values of true positives, true negatives, and MCC are virtually identical, while their standard deviations across all 10 runs all are smaller than one percent (see table below). In addition to its small absolute size, this result for MCC (0.0098) compares reasonably well against the approximate standard error of the MCC estimator (0.0011) based on its estimated 95% confidence interval (see Itaya et al. (2024), eq. (7)).

**TABLE E: %Rightly Identified in Each Cell**
**A. 12 Cells with Directional Relationships, B. 33 Cells with No Relationships**

| | I | C | E | G | L | S | M | J | D | Z |
|---|---|---|---|---|---|---|---|---|---|---|
| I | | | | | | | | | | |
| C | | | | | | | | | | |
| E | 85.4% | 85.2% | | | | | | | | |
| G | | 98.1% | | | | | | | | |
| L | | | | | | | | | | |
| S | | | 71.7% | | 72.4% | | | | | |
| M | | | 72.1% | | | | | | | |
| J | | | | | | | | | | |
| D | | | 12.5% | 60.0% | 55.4% | | 30.6% | 70.3% | | |
| Z | | | | | | | | | 95.9% | |

| | I | C | E | G | L | S | M | J | D | Z |
|---|---|---|---|---|---|---|---|---|---|---|
| I | | | | | | | | | | |
| C | 98.3% | | | | | | | | | |
| E | | | | | | | | | | |
| G | 95.9% | | 98.0% | | | | | | | |
| L | 96.2% | 96.3% | 97.2% | 98.5% | | | | | | |
| S | 99.1% | 99.0% | | 97.0% | | | | | | |
| M | 99.8% | 99.8% | | 97.6% | 98.2% | 99.0% | | | | |
| J | 95.8% | 96.2% | 98.0% | 96.9% | 98.7% | 98.0% | 98.9% | | | |
| D | 98.1% | 99.3% | | | | 99.3% | | | | |
| Z | 96.7% | 97.9% | 98.3% | 98.4% | 98.3% | 97.7% | 99.3% | 99.1% | | |

| | True Positives | True Negatives | MCC |
|---|---|---|---|
| **Original** | 67.46% | 98.01% | 73.15% |
| **Mean: 10 Column Orders** | 67.72% | 98.05% | 73.43% |
| **Stdev: 10 Column Orders** | 0.88% | 0.17% | 0.98% |

For settings where the conditions listed above (e.g. extreme sparsity, nearly perfect collinearity, etc.) are not uncommon, this would not be an unreasonable check to perform when implementing ADD. But empirically here, even under this study's reasonably challenging data conditions, the effects of column-ordering are not material, and it is worth noting that two other studies confirm this empirical result of non-materiality vis- à-vis column ordering (see Stollenwerk (2026) and Chiriac and Voev (2010)).

**Graph 5: Distribution of #Rightly Identified Directional Effects: 12 Pairs with Effects (Fully Exploratory Analysis, No Priors)**

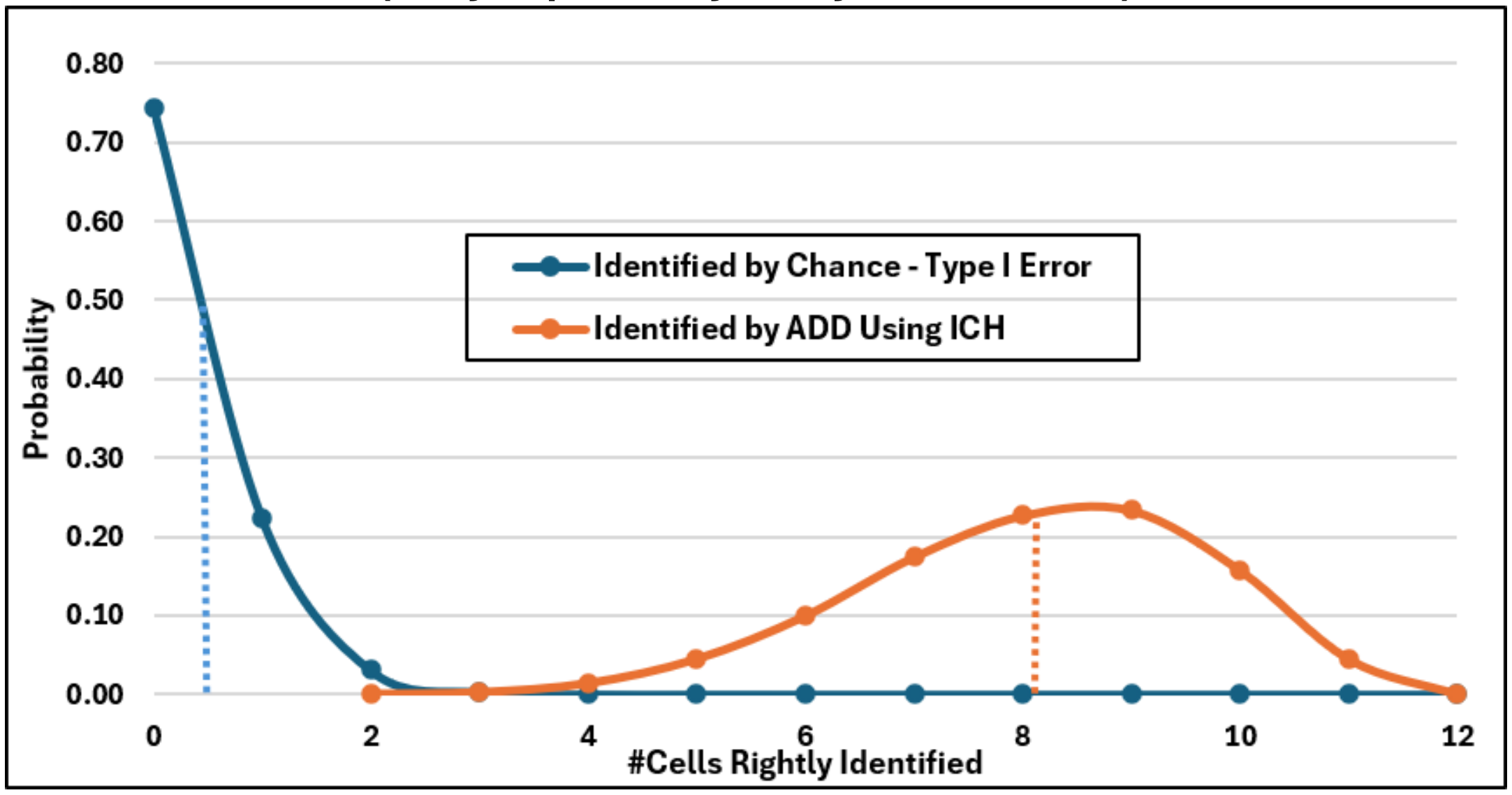

**Graph 6: Distribution of #Rightly Identified Directional Effects: 12 Pairs with Effects (Perfect Priors on which 12 Pairs Have SOME Effect, but not the Direction)**

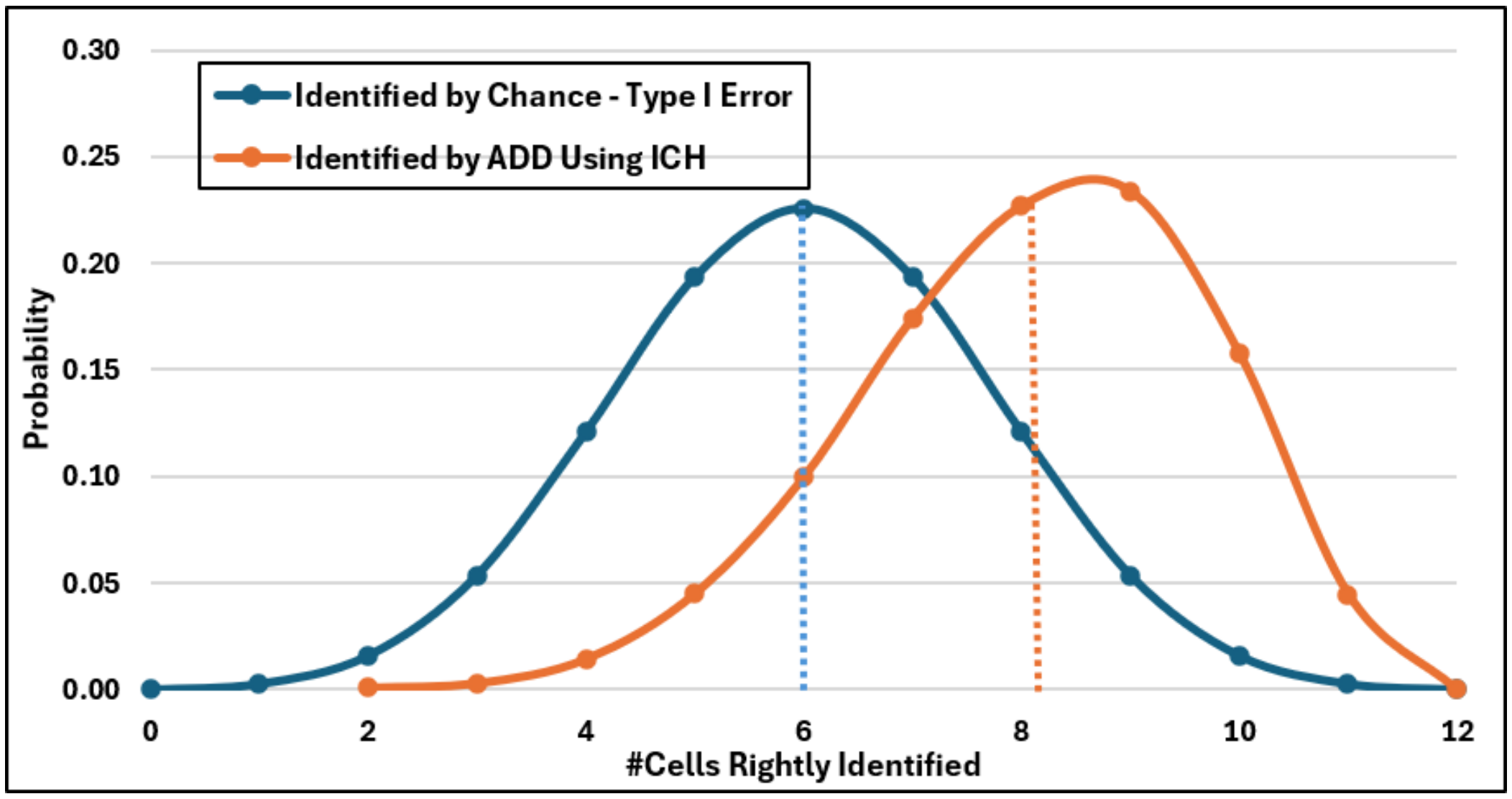

**TABLE F: ADD's DAG Recovery vs. No Model**

All 45 Pairs, Fully Exploratory Analysis, No Priors

| no model* | | ADD-ICH | | |
|---|---|---|---|---|
| # right | % right | # right | % right | p-value* |
| **31.66** | **70.4%** | **40.44** | **89.9%** | **<0.00001** |

*Binomial vector: p12=α_adj*(1-α_adj), p33=p=[1 - α_adj]^2

*p-value: permutation test, nsims=100k, n1=n2=45

12 Pairs with Effects, Fully Exploratory Analysis, No Priors

| no model** | | ADD-ICH | | |
|---|---|---|---|---|
| # right | % right | # right | % right | p-value** |
| **0.29** | **2.4%** | **8.09** | **67.5%** | **<0.00001** |

**Binomial distribution, p12=α_adj*(1-α_adj)

**p-value: permutation test, nsims=100k, n1=n2=12

12 Pairs with Effects, Perfect Priors on Right Pairs but not Direction of Effect

| no model~ | | ADD-ICH | | |
|---|---|---|---|---|
| # right | % right | # right | % right | p-value~ |
| **6.00** | **50.0%** | **8.09** | **67.5%** | **0.01174** |

~Binomial distribution, p12=0.5

~p-value: permutation test, nsims=100k, n1=n2=12

33 Pairs with No Effect, Fully Exploratory Analysis, No Priors

| no model~~ | | ADD-ICH | |
|---|---|---|---|
| # right | % right | # right | % right |
| **31.37** | **95.1%** | **32.34** | **98.0%** |

~~Binomial distribution, p33=[1 - α_adj]^2

This is a preliminary and relatively narrow empirical study, but its results are promising for ADD's general application to DAG recovery for causal modeling. While ADD is not designed to provide effect sizes or counterfactual outcomes, it appears to be able to effectively identify causal structure via DAG recovery, under the assumption of direct adjacency, with reasonable power.

Importantly, ADD maintains the flexibility to use any (positive definite) directional dependence measure for this problem of DAG recovery, as befits the data conditions of a particular setting, which in itself gives it advantages over many competitors. While ADD does inherit the strengths and weaknesses of the (directional) dependence measure used, and these will vary notably based on the data types being evaluated, careful scrutiny of the data can turn this potential drawback into a strong comparative advantage, permitting the opportunistic selection of dependence measures to increase power over less flexible causal algorithms. In settings where data conditions are less well defined, more difficult to estimate, and/or subject to more frequent regime changes, one way to maintain and possibly magnify this advantage presents itself in the earlier literature as 'stacking' (see Breiman (1996), Wolpert (1992), Odegua (2019), and Barton and Lennox (2022)). To use it here we run ADD multiple times on the same input data but using different dependence measures, and then 'stack' these multiple results. The goal is to cover, in a single implementation, a broad(er) range of possible data conditions with greater power on average. As long as multiplicity is properly accounted for, this may be a very powerful option under these challenging but realistic conditions, and one we will explore in the follow-up empirical study.

Additionally, ADD's simultaneous estimation of all pairwise relationships, which properly and automatically enforces positive definiteness, may provide greater coherence in its edge calls compared to other causal algorithms, especially those that call edges sequentially. This has been a serious obstacle plaguing many of the widely used DAG recovery algorithms to date (see Hulse et al. (2025) and Faltenbacher et al. (2026)), so this is a nontrivial potential advantage of ADD. Finally, the fact that ADD requires only two matrix estimations, and two fast simulations to establish confidence intervals under independence, means that it scales efficiently with the size of the DAG.

All of these actual and potential advantages will be thoroughly explored and tested in a more extensive, follow-up empirical study that compares ADD to competitor algorithms. The study will include different sample sizes, different data types (e.g. both stationary and non-stationary), varying signal-to-noise ratios, different directional dependence measures, different DAGs, and different pruning algorithms, so it will be quite comprehensive. But first implementing this narrower, yet reasonably challenging study was an efficient, useful, and large first step on this path of empirical testing and validation.

## 5. Conclusions and Next Steps

In ADD, we develop an original approach to causal discovery via DAG recovery: ADD simultaneously identifies directional dependence using dual orderings in the angles space of directional dependence measures. Our preliminary empirical study provides promising results (in the sense of direct adjacency) under challenging data conditions similar to those encountered in quantitative finance settings (i.e. nonlinearity, asymmetry, and heavy-tailedness). This indicates ADD's potential utility for feature selection and model building in this setting. Additional potential advantages include increased coherence, due to simultaneous edge-calling within a positive definite space; increased power, due to the flexible capability to use any directional dependence measure and thus, opportunistically adapt to different or varying data conditions; and increased speed/scalability, due to the need for only two matrix estimations, and two fast simulations to define empirical confidence bounds under independence, regardless of the size of the DAG space. A more extensive follow-up study will test these by directly benchmarking performance against competing causal discovery algorithms.

We end on a somewhat cautionary note, however, recognizing that this dive into causality via DAG recovery begs the bigger and valid question of whether DAGs can be used reliably within "self-referencing open systems like capital markets" to begin with (Polakow et al., 2023). This applies to other settings as well, because DAGs have strong limitations generally, as cited herein, cautioned by Dawid (2009) and MacKinnon and Lamp (2021), and demonstrated in recent research by Hulse et al. (2025), Padh et al.

(2025), Kaiser and Sipos (2022), Faltenbacher et al. (2026), Gong et al. (2024), and De Lara (2024).[42] But alongside the new potential benefits of ADD, progress continues to be made here. Recent innovative work on geometric causal models is providing both broader and more flexible structural frameworks for causal discovery (see Weinstein and Blei (2026) and Usta and Bollt (2024), respectively). And regarding the issue of temporality specifically, Reisach et al. (2026) persuasively contend that a major limitation of "nontemporal" causal DAGs is that they fail to explicitly incorporate the notion of time into the DAG, thus arguably obstructing justification of the acyclicity assumption; they address this by creating *composite* causal variables that refer to quantities at one or multiple time points. Similarly, the work of Cüppers et al. (2024) explicitly enables dynamic delays in their causal framework using a Minimum Description Length to add edges in their discovery algorithm in topological order. And the recent work of Rodriguez Dominguez (2023, 2025a, 2025c) addresses the challenging issue of temporality by embedding association-based dependence measures directly within time-dependent causal frameworks. With a focus on this latter work, we will explore integration between ADD and these methodologies to directly address temporal issues in the follow-up study.

[42] From Polakow et al. (2023): "The clarion call for causal reduction in the study of capital markets is intensifying. However, in self-referencing and open systems such as capital markets, the idea of unidirectional causation (if applicable) may be limiting at best, and unstable or fallacious at worst." From Gong et al. (2024): "... potential outcomes (PO) and structural causal models (SCMs) stand as the predominant frameworks. However, these frameworks face notable challenges in practically modeling counterfactuals ... we identify an inherent model capacity limitation, termed as the 'degenerative counterfactual problem', emerging from the consistency rule that is the cornerstone of both frameworks." And from De Lara (2024): "Most of the literature on causality considers the structural framework of Pearl and the potential-outcomes framework of Neyman and Rubin to be formally equivalent, and therefore interchangeably uses the do-notation and the potential-outcome subscript notation to write counterfactual outcomes. In this paper, we ... prove that structural counterfactual outcomes and potential outcomes do not coincide in general – not even in law." See Opdyke (2024b) for a more complete review of this literature.

## 7. APPENDIX A1: Translations Between Dependence Measure Matrices & Angles Matrices

Scenario 1 (S1): Identity Matrix

**Dependence Measure Matrix**

| | | | | |
|---|---|---|---|---|
| 1 | 0 | 0 | 0 | 0 |
| 0 | 1 | 0 | 0 | 0 |
| 0 | 0 | 1 | 0 | 0 |
| 0 | 0 | 0 | 1 | 0 |
| 0 | 0 | 0 | 0 | 1 |

**Angles Matrix**

| | | | | |
|---|---|---|---|---|
| | | | | |
| π/2 | | | | |
| π/2 | π/2 | | | |
| π/2 | π/2 | π/2 | | |
| π/2 | π/2 | π/2 | π/2 | |

Scenario 2 (S2): Block Matrix

**Dependence Measure Matrix**

| | | | | |
|---|---|---|---|---|
| 1 | -0.2 | -0.2 | 0.3 | 0.3 |
| -0.2 | 1 | -0.2 | 0.3 | 0.3 |
| -0.2 | -0.2 | 1 | 0.3 | 0.3 |
| 0.3 | 0.3 | 0.3 | 1 | 0.6 |
| 0.3 | 0.3 | 0.3 | 0.6 | 1 |

**Angles Matrix**

| | | | | |
|---|---|---|---|---|
| | | | | |
| 1.772 | | | | |
| 1.772 | 1.823 | | | |
| 1.266 | 1.175 | 1.002 | | |
| 1.266 | 1.175 | 1.002 | 1.295 | |

Cell # Matrix for Angles PDFs below in Appendix A3:

**Cell # Matrix for PDFs**

| | | | | |
|---|---|---|---|---|
| | | | | |
| 7 | | | | |
| 8 | 4 | | | |
| 9 | 5 | 2 | | |
| 10 | 6 | 3 | 1 | |

APPENDIX A Comments:

1. As expected, note that the identity (independence) matrix yields symmetric angles PDFs centered on π/2. Also as expected, the corresponding empirical spectral distribution aligns with that of Marchenko Pastur (see Marchenko and Pastur (1967)).

2. As expected, the spectral distribution of Pearson's rho is more variable than that of Kendall's tau, as it is generally more sensitive to outlying observations.

3. Note that the characteristics of the spectral distributions compared to those of the angles distributions, i.e. more asymmetry, multi-modality, and long, essentially unbounded tails, make their estimation more challenging and less robust than that of the individual angles distributions. See Opdyke (2022, 2026) for more on this topic.

4. Note the larger scale of the y-axes for the improved Chatterjee PDFs.

## 7. APPENDIX A2: Spectral Distributions Under Five Measure/Scenario Combinations

**Scenarios (n=252, Marchenko Pastur (Independence) Added as Referential Baseline)**

**S1**: Measure = Pearson's Rho; Dependence Structure = Identity Matrix; DGM = Multivariate Gaussian

**S2**: Measure = Pearson's Rho; Dependence Structure = Block Matrix; DGM = Multivariate Gaussian

**S2A**: Measure = Pearson's Rho; Dependence Structure = Block Matrix;
DGM = Marginal Distributions with Different Asymmetry, Heavy-Tailedness, Serial-Correlation, and Non-stationarity (via Church's (2012) asymmetric student's t copula with varying degrees of freedom: serial correlation and non-stationarity imposed ex post, with subsequent rescaling of the first two moments)

**S2B**: Measure = Kendall's tau; Dependence Structure = Block Matrix; DGM = Same as S2A

**S2C**: Measure = Improved Chatterjee's; Dependence Structure = Block Matrix; DGM = Same as S2A

**Spectral Distributions: S1, S2, S2A, S2B, S2C**

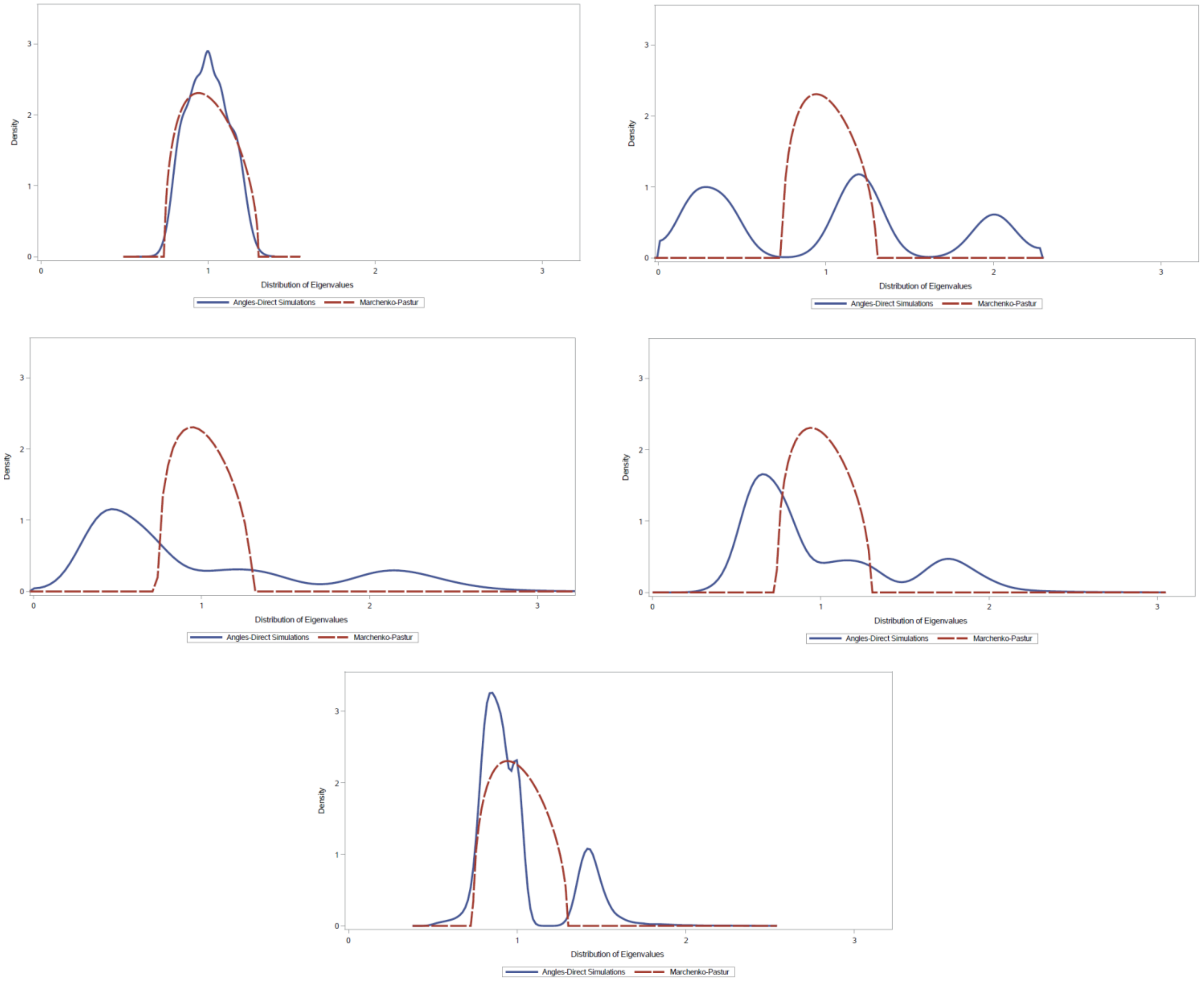

## 7. APPENDIX A3: Scenario S1 Angles PDFs: n=252, NSims=10k, Cell # Per Appendix A1 (1, 2, ..., 10)

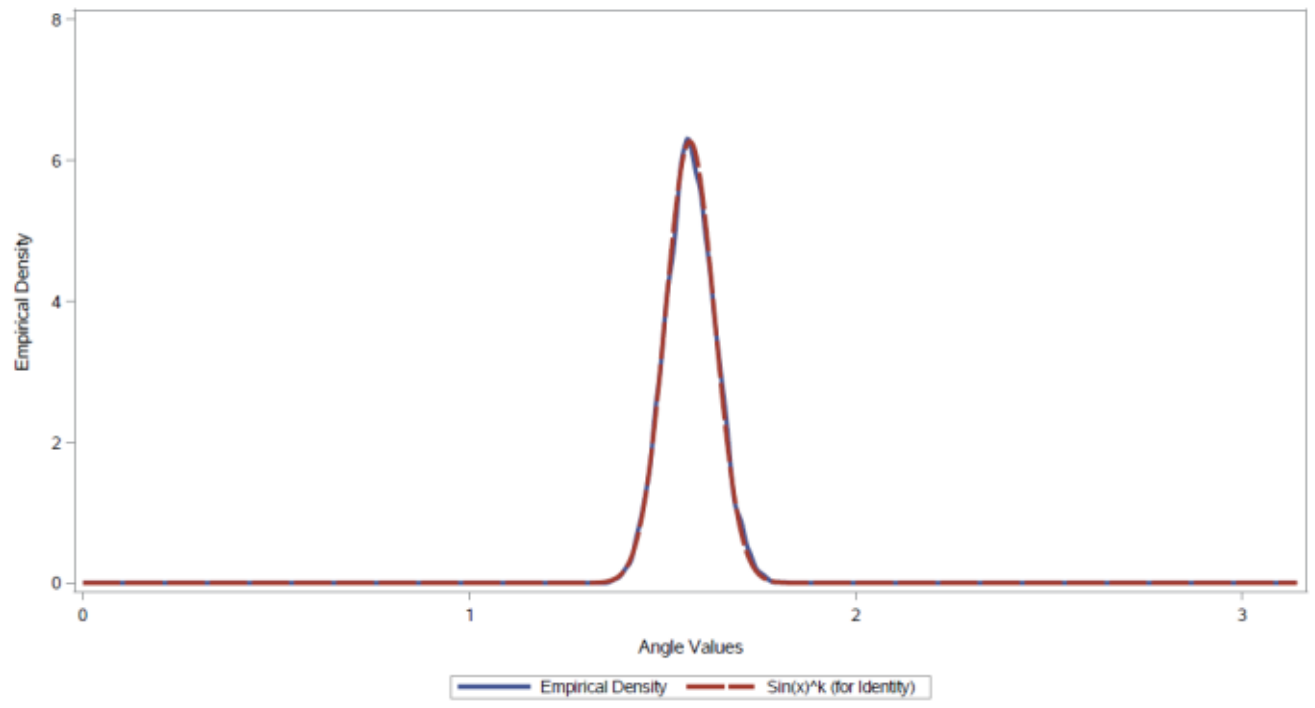

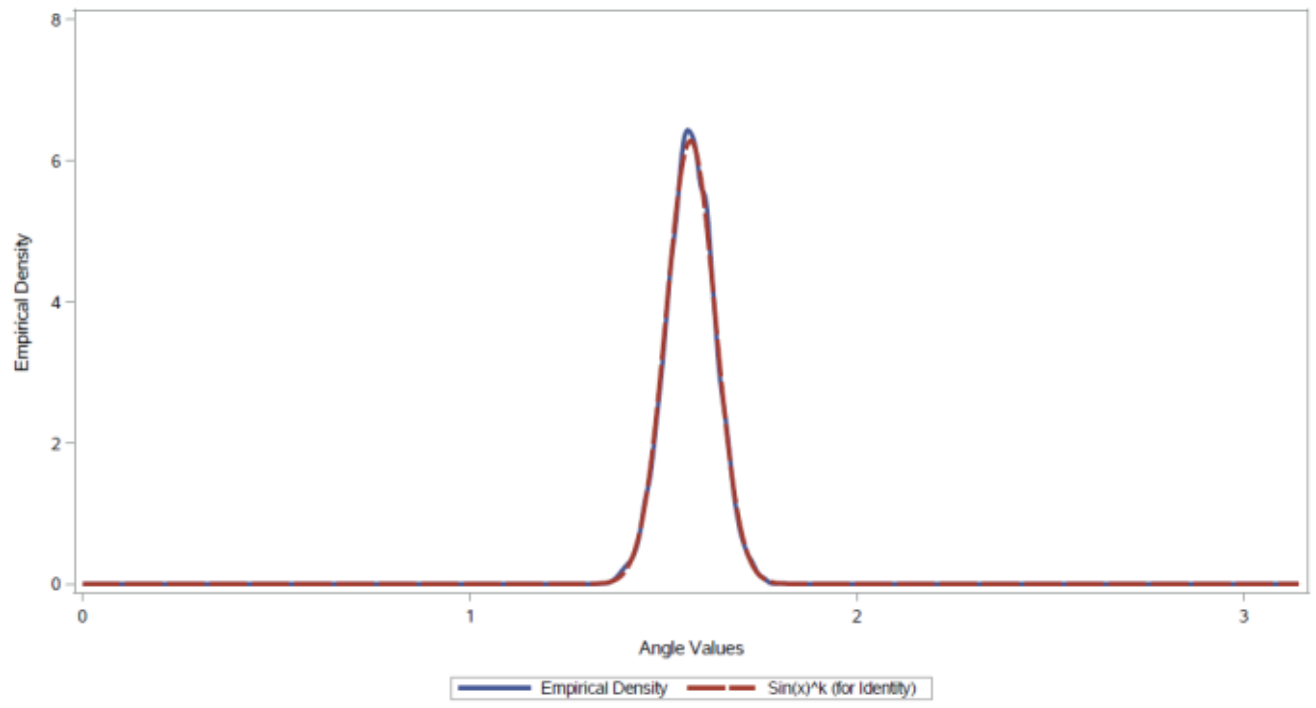

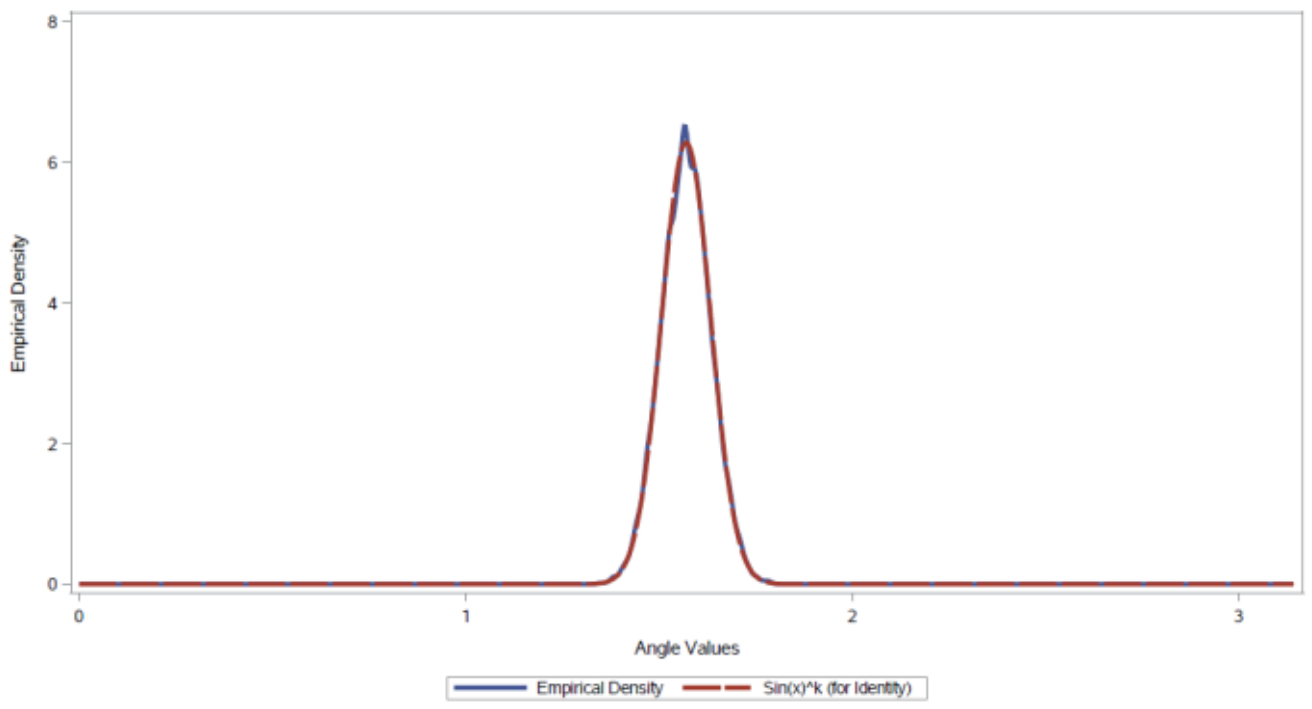

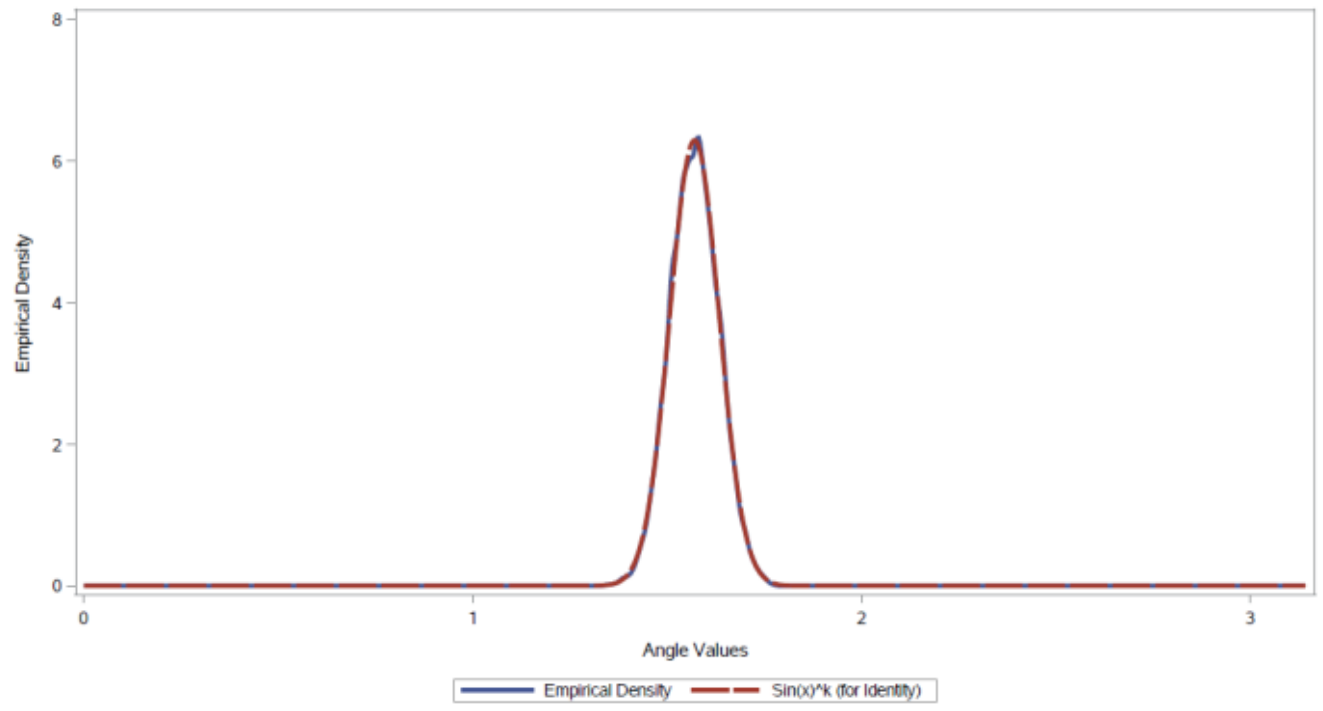

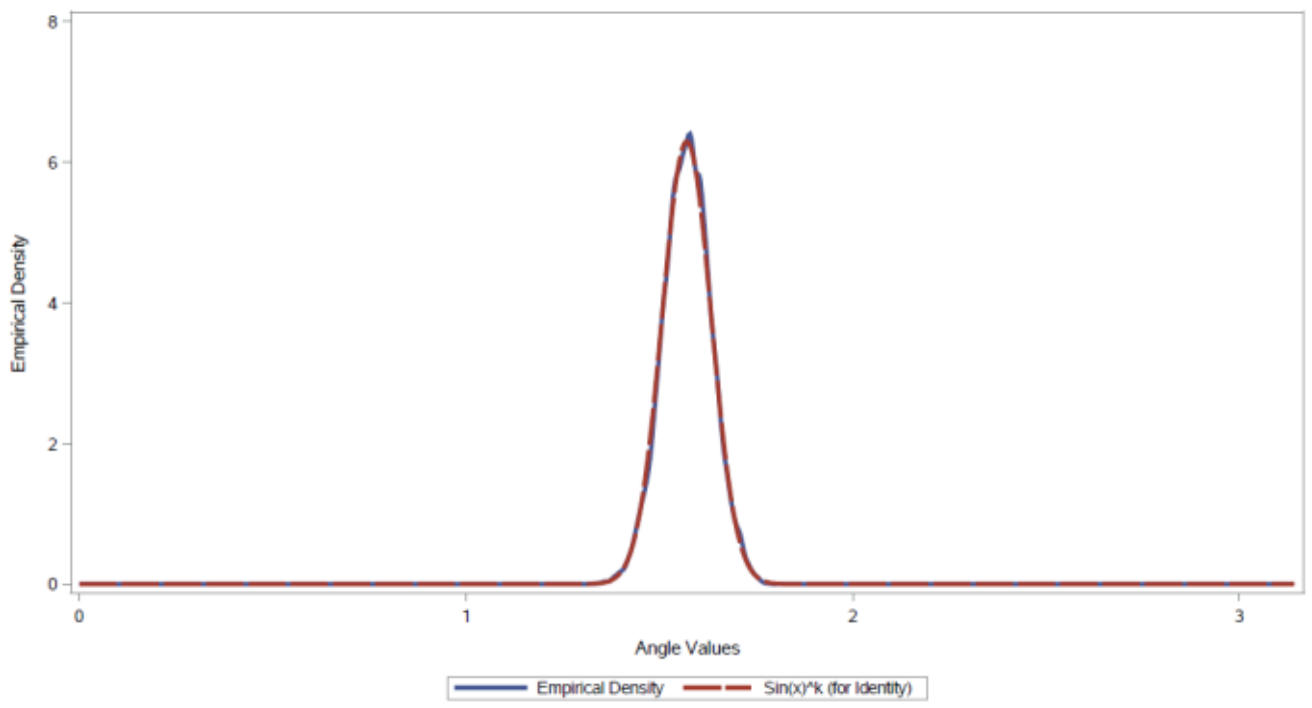

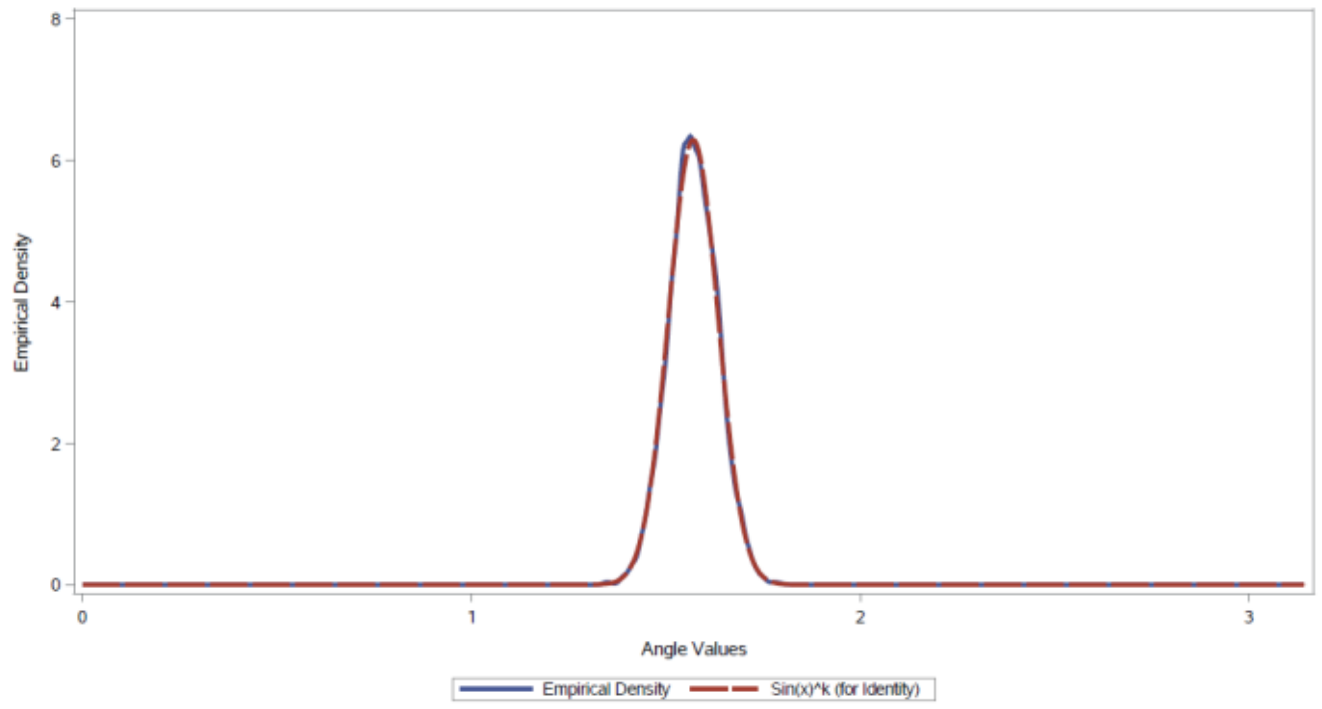

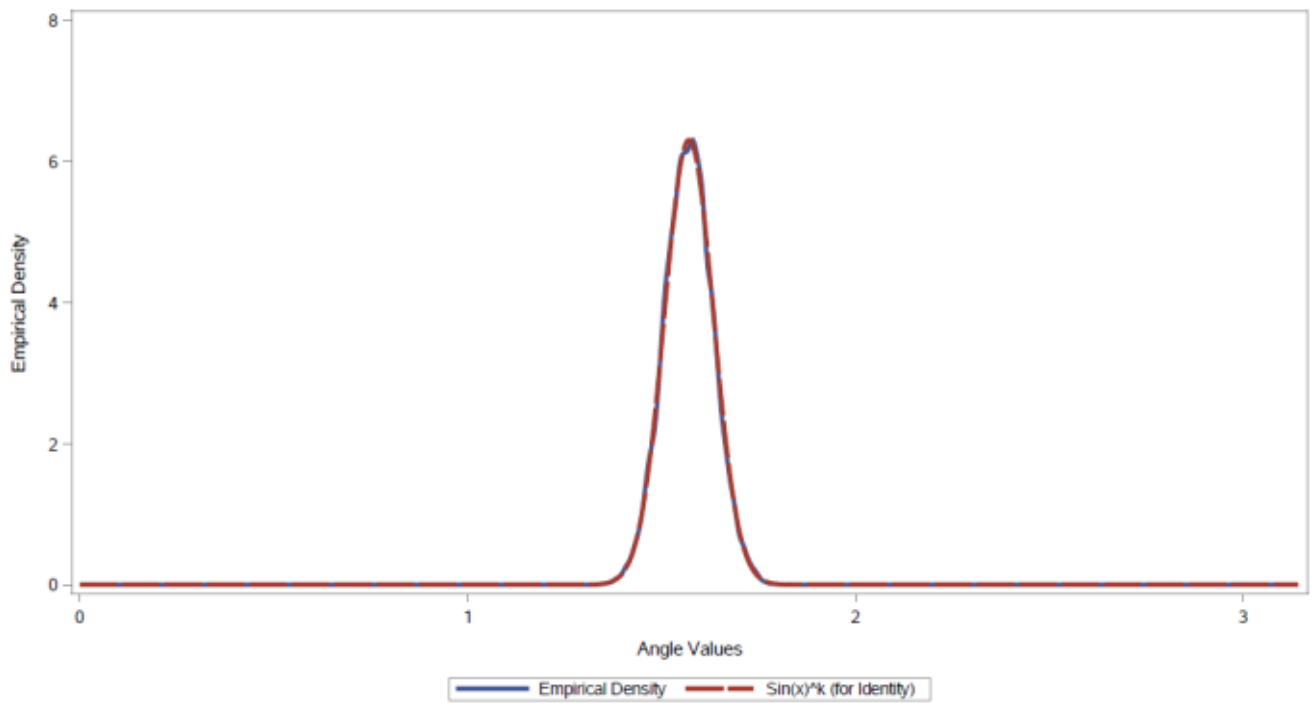

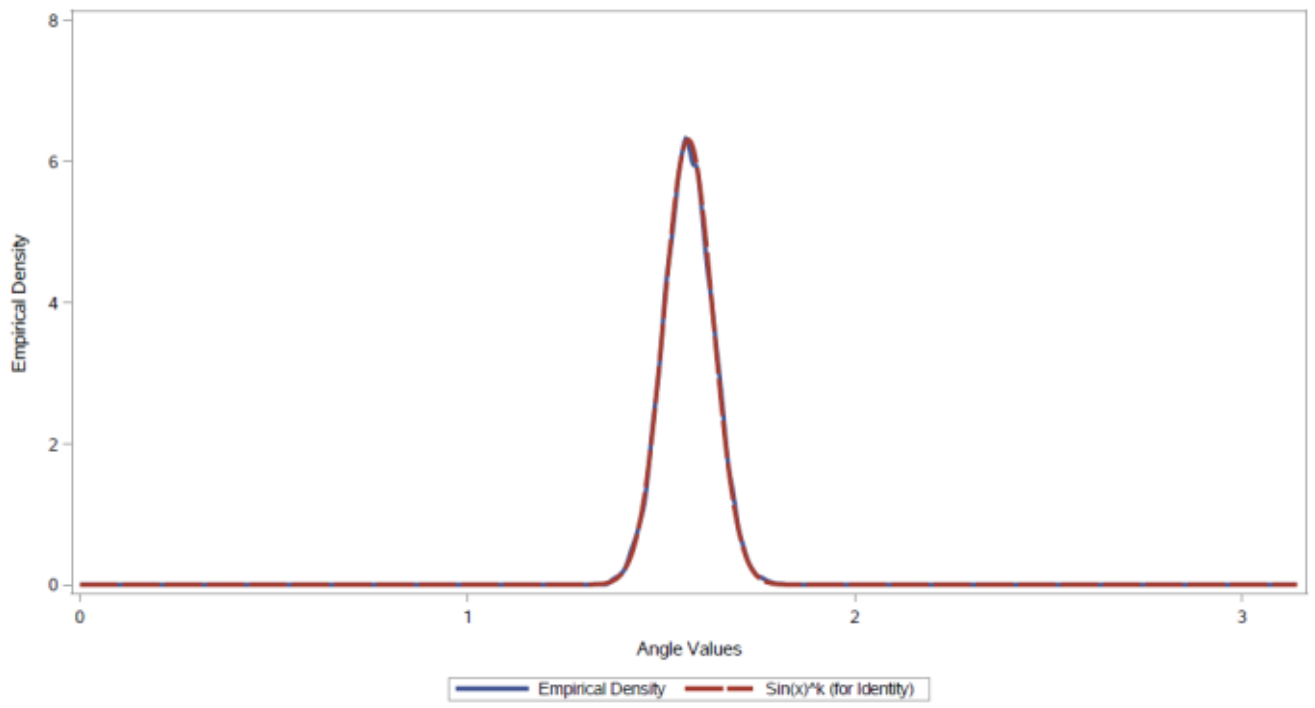

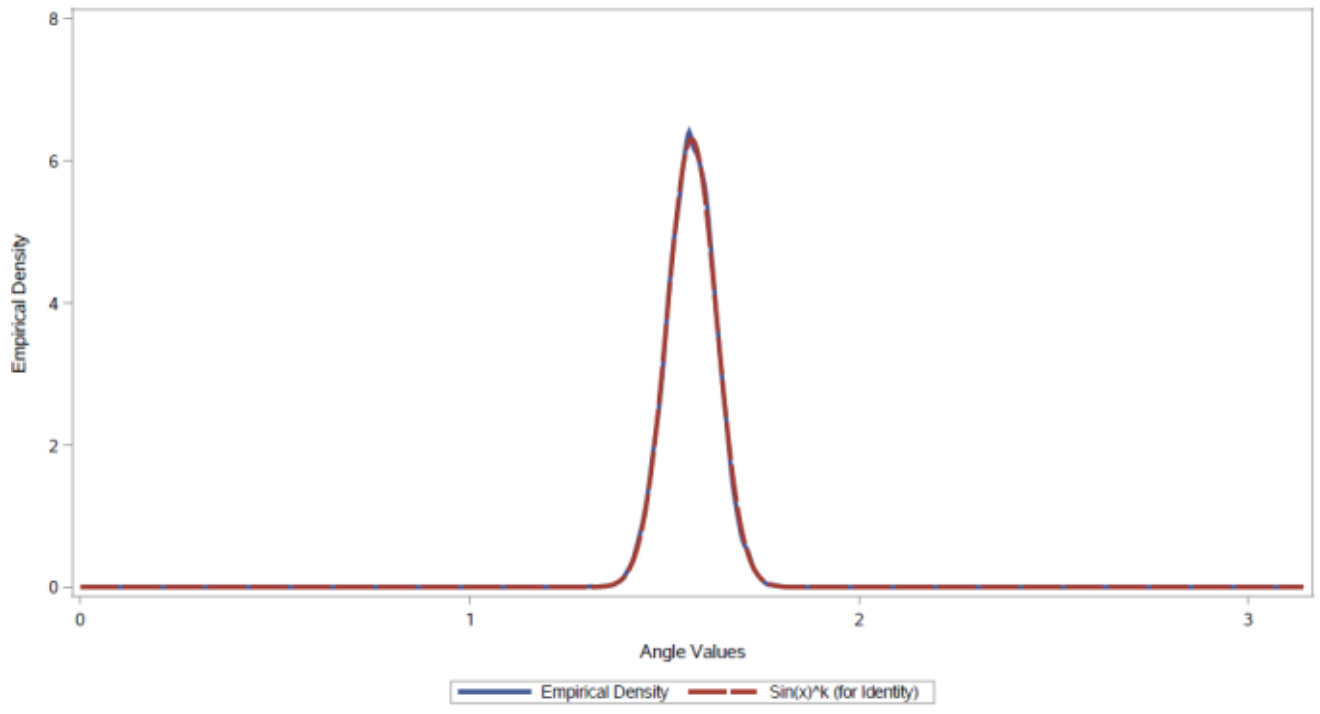

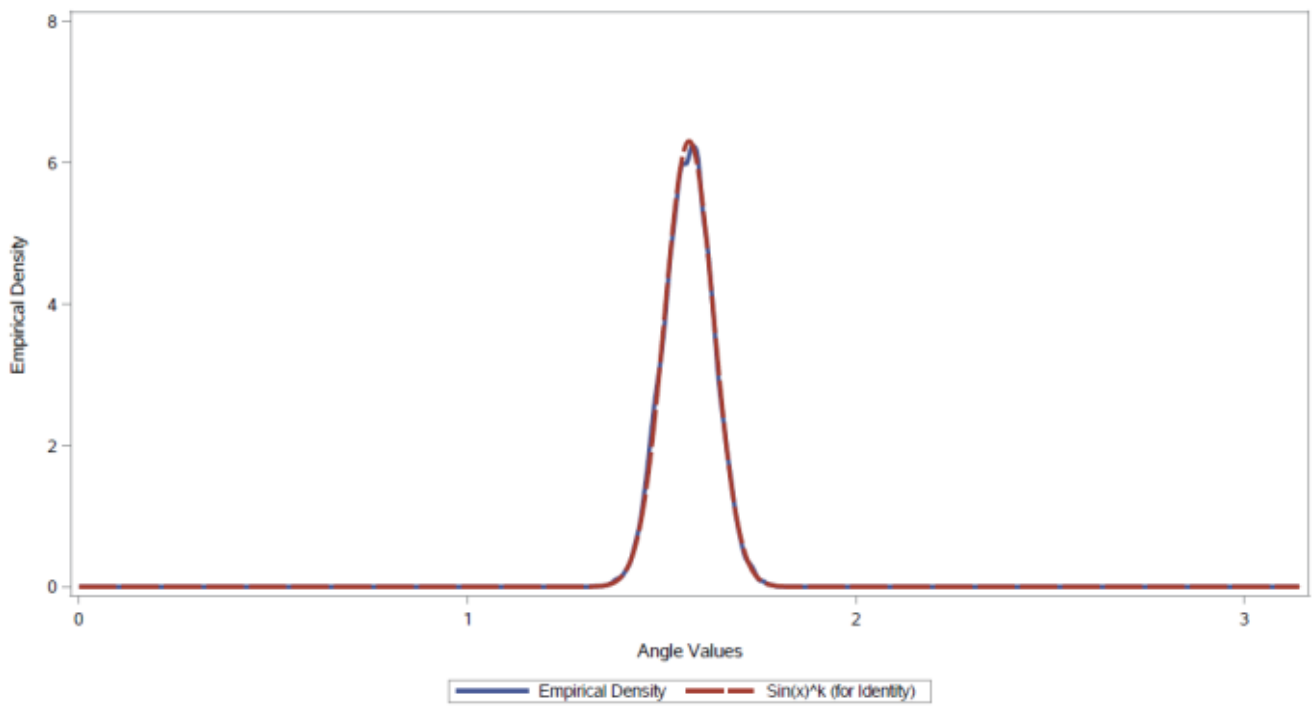

**7. APPENDIX A3: Scenario S2 Angles PDFs:** n=252, NSims=10k, Cell # Per Appendix A1 (1, 2, ..., 10)

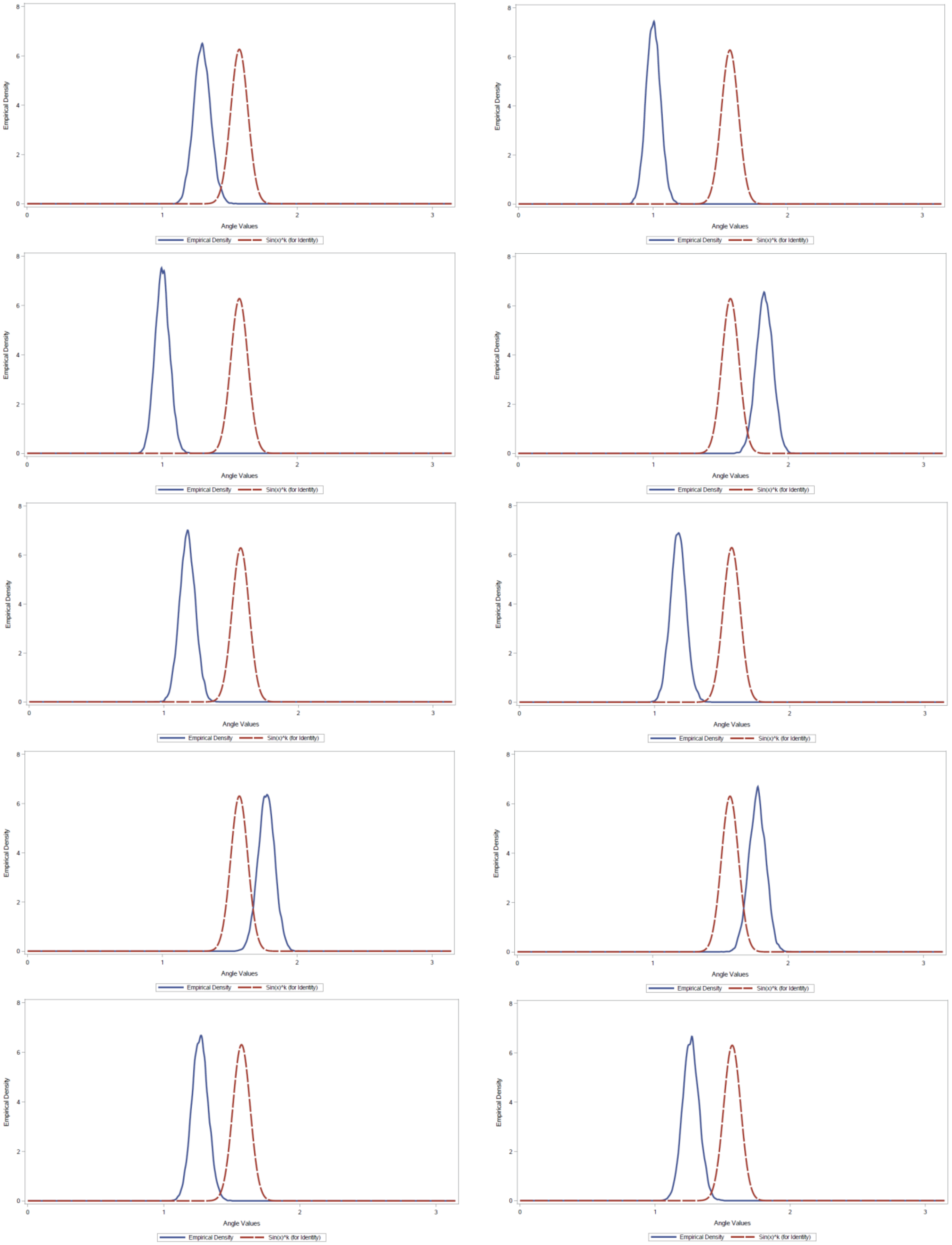

**7. APPENDIX A3: Scenario S2A Angles PDFs:** n=252, NSims=10k, Cell # Per Appendix A1 (1, 2, ..., 10)

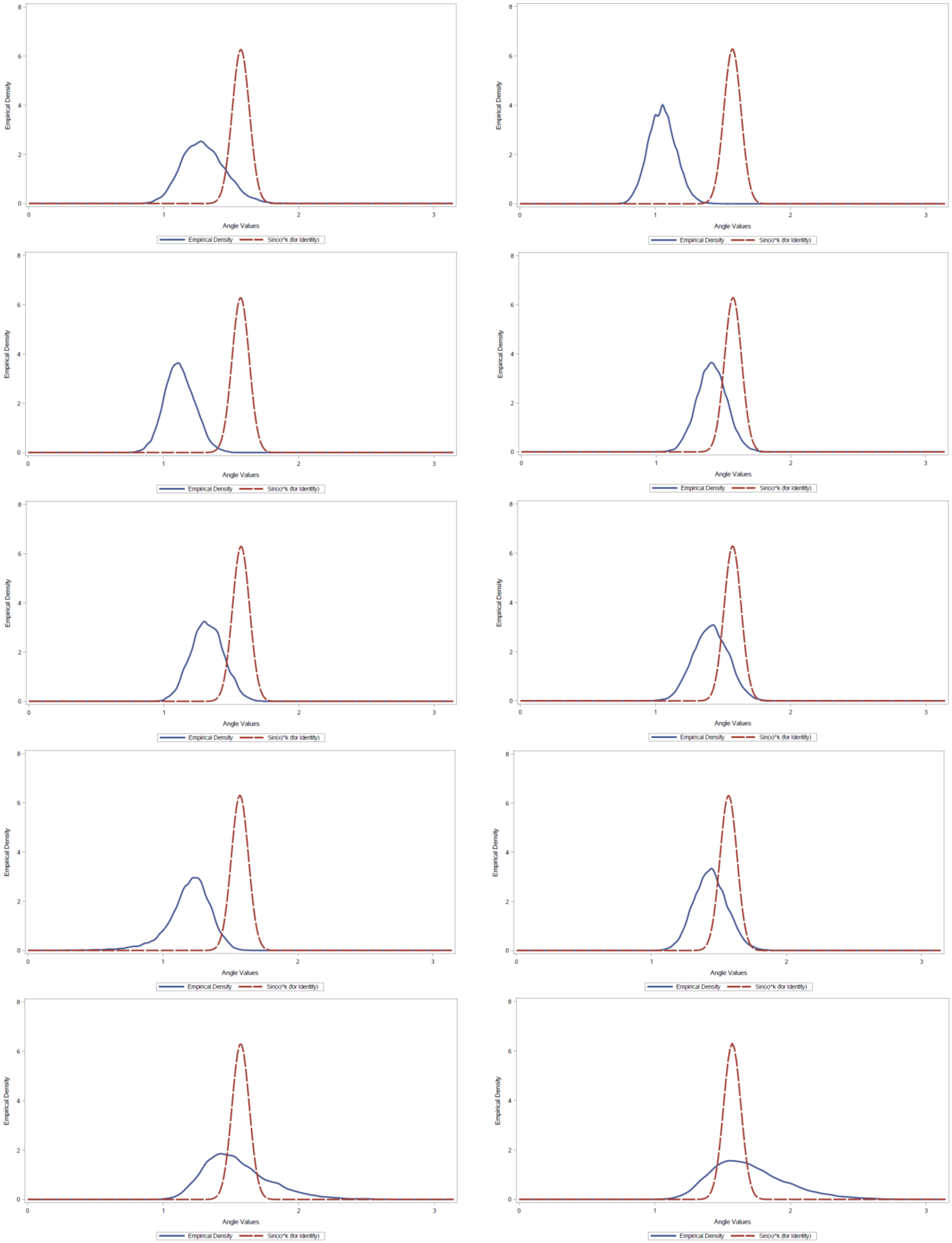

## 7. APPENDIX A3: Scenario S2B Angles PDFs: n=252, NSims=10k, Cell # Per Appendix A1 (1, 2, ..., 10)

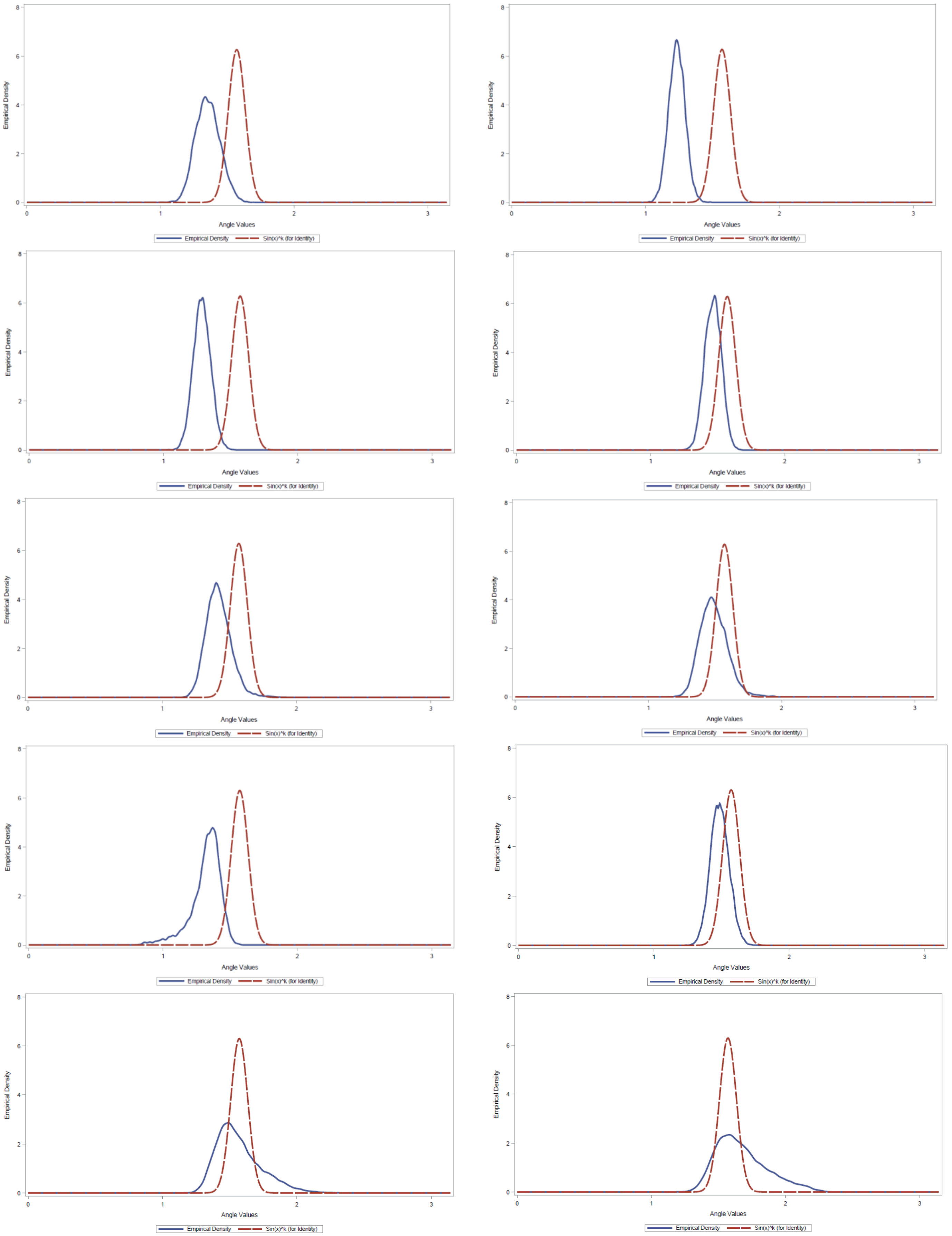

## 7. APPENDIX A3: Scenario S2C Angles PDFs: n=252, NSims=10k, Cell # Per Appendix A1 (1, 2, ..., 10)

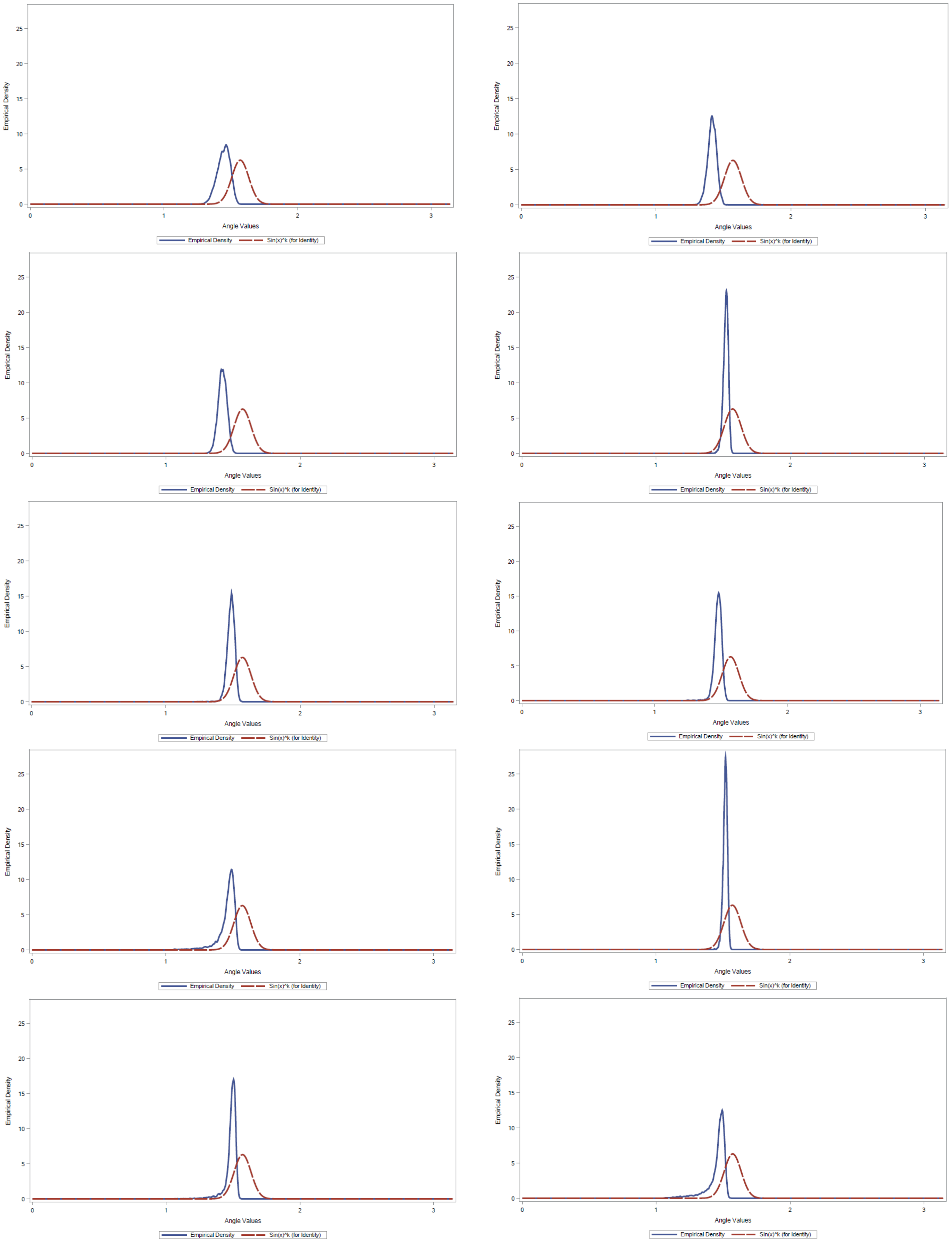

**7. Appendix B: Efficient Coding of Improved Chatterjee Correlation of Xia et al. (2025)**

## SAS/IML code (v9.4): Improved Chatterjee

```
*** INPUTS: data4chat is raw input returns data, reverse is a flag to reverse column order;
*** OUTPUT ichat is the improved Chatterjee matrix;

  start corr_ichat(data4chat,reverse);
    dim = dimension(data4chat);
    ncols=dim[1,2];
    ncols_m1=ncols-1;
    nrows=dim[1,1];
    nrows_m1=nrows-1;
    ncells=ncols**2;
    ichat=J(ncols,ncols,.);
    ichat[do(1,ncells,ncols+1)]=1;
    if reverse=1 then data4chat=data4chat[,do(ncols,1,-1)];
    if nrows>ncols then do;
      do i=1 to ncols_m1;
        data4chat2=data4chat[,i:ncols];
        call sort(data4chat2);
        do j=i+1 to ncols;
          k=j-i+1;
          rnks = rank(data4chat2[,k]);
          cumsum=0;
          cumwgts=0;
          do q=1 to nrows_m1;
            cumsum=cumsum+sum(abs(dif(rnks,q)/q));
            cumwgts=cumwgts+(nrows-q)/q;
          end;
          ichat[i,j]=1-cumsum/(cumwgts*(nrows+1)/3);
          ichat[j,i]=ichat[i,j];
        end;
      end;
   end;
   else do;
     print "Data provided to corr_ichat subroutine must be full rank.";
   end;
   return(ichat);
  finish;
```

Note that the "improved Chatterjee correlation" of Xia et al. (2025) typically would be coded with a nested loop, but this can be avoided by calculating the entire equation <u>diagonally</u>, thus increasing speed by a factor of p (the dimension of the matrix). This can save considerable amounts of real runtime when matrices are large(r) (e.g. p≥100) AND these matrices must be calculated many times in many simulations (e.g. Nsim ≥10,000).

## 7. Appendix C: Assumptions Required for Valid Causal Interpretation

**A.** Below we define the assumptions under which the directional associations identified by ADD can be interpreted causally, and the causal meaning of edges inferred by ADD in a DAG.

**Preliminaries**: Structural Causal Model (SCM) and do-Semantics (see Pearl, 2009)

Let $V = \{X_1, \ldots, X_p\}$ be observed variables generated by an acyclic Structural Causal Model (SCM):

$$X_j = f_j\left(X_{Pa(j)}, U_j\right) \text{ for } j = 1, \ldots, p$$

Here Pa(j) are the parents of $X_j$ in a DAG "G," $U_1, \ldots, U_p$ are jointly independent exogenous noises, and each $f_j(\ )$ is a (possibly nonlinear, nonparametric) measurable function. The graph G encodes the causal Markov property and admits do-interventions: for any variable X, the operation do(X:=x) replaces its structural equation with the constant x, severing incoming edges into X.

Causal effect: X has a (possibly context-dependent) causal effect on Y if and only if there exist x ≠ x' such that P(Y | do(X=x)) ≠ P(Y | do(X=x')).

**Assumptions for Causal Interpretation**:

(A1) Acyclicity: The true causal graph over V is a DAG.

(A2) Causal Sufficiency (no latent confounders among V): The exogenous noises $U_j$ are mutually independent; any common causes of variables in V are included in V. Relaxation: if (A2) is uncertain, interpret edges as ancestral (existence of a directed path) rather than necessarily adjacent (direct).

(A3) Markov + (qualified) faithfulness: The observed joint distribution is Markov with respect to G and violations of direction-revealing asymmetries are non-degenerate (i.e. pathological cancellations are excluded).

(A4) i.i.d. sampling (or weak stationarity for time-series): Samples are independent draws from the same distribution (future research plans to apply ADD, for time-series extensions, to innovations or appropriate lags).

(A5) Measurement reliability: Variables are measured without systematic differential error that inverts directional asymmetries.

(A6) Conditioning set for direct edges: When inferring adjacency (direct edges), directional dependence is evaluated conditionally on V\{X,Y}. If only marginal directional dependence is used, interpret edge calls as directed reachability (existence of some directed path).

(A7) Monotone link from mechanism to asymmetry: For the chosen directional measure *M* (in the empirical study above, this is ICH – the improved Chatterjee's correlation of Xia et al. (2025)), if X ∈ Pa(Y) and Y ∉ Pa(X), then asymptotically E[ $M$(X→Y) ] > E[ $M$(Y→X) ], with separation increasing in sample size n under the SCM.

Note that this assumption plays a fundamental role in linking the structural causal model to observable directional asymmetries. It requires that the directional dependence functional $M$ increase in expectation in the true causal direction. This monotone link between mechanism and measurable asymmetry is the critical identifiability condition for ADD-based DAG recovery. In practice, validation of (A7) proceeds through a simulation across a range of nonlinear, non-Gaussian, and/or heteroskedastic settings, ensuring that E[ $M$(X→Y) ] > E[ $M$(Y→X) ] consistently emerges under the data-generating process.

Among the seven assumptions, only (A7) must be verified empirically.

**Edges in ADD**:

Let $M$ be a directional dependence functional and $\hat{M}$ its sample estimate. ADD maps the full matrix of $\hat{M}$ values to angles, performs finite-sample joint inference under positive-definiteness, and returns p-values for directed hypotheses.

Adjacency claim (direct edge): Under (A1)–(A7) and using conditional directional dependence $M$(X→Y | V\{X,Y}): declare a directed edge X→Y if statistically significant(X→Y) and NOT statistically significant(Y→X) at level α (the one-direction-only rule). This is interpreted as X ∈ Pa(Y).

Ancestral/path claim (directed reachability): If only marginal $M$ is used, interpret the same rule as: there exists a directed path X ⇝ Y (not necessarily adjacent); if (A2) is violated, reduce the claim to directional dependence without causal interpretation.

No-edge claim: If neither direction is statistically significant, refrain from a causal claim.

Cycle handling: If both directional tests are significant, yielding a 2-cycle, pruning enforces acyclicity by selecting the edge/direction that has a larger p-value and replacing its angle value with a value just below what would indicate statistical significance at the α threshold. The same is done when longer cycles are encountered. The specific algorithm for the latter case that was implemented in the above empirical study used Kahn's (1962) topological sorting algorithm to identify cycles, and then i. selected those nodes/variables with the fewest edges; ii. then among those, selected the one with the largest sum of p-values across its related edges (i.e. the weakest cumulative signals/edges); and iii. finally, selected the particular edge from among those that had the largest p-value (weakest signal/edge). This process was repeated until no cycles remained. In the above empirical study, no pruned matrices became non-positive definite, which is not surprising given that the pruning adjustments place the matrix further within the positive definite convex cone (still, positive definiteness always was empirically verified). But even if positive definiteness had been occasionally violated, non-linear constrained optimization is relatively straightforward to implement here. While this pruning algorithm is not multivariate optimized, it is fast and straightforward, and was effective in the empirical study herein. A follow-up study will explore optimizing this approach, possibly by minimizing an aggregate angle-space loss function under positive definite constraints.

Under (A1)–(A7), the one-direction-only ADD decision rule using conditional *M* is pointwise sound for adjacency: if the rule declares X→Y, then X ∈ Pa(Y) with probability → 1 as n → ∞.

By Markov + faithfulness (A3) and SCM acyclicity (A1), conditioning on V\{X,Y} blocks all non-causal paths between X and Y. If X ∈ Pa(Y), then by (A7) the population asymmetry satisfies *M*(X→Y) > *M*(Y→X); ADD's finite-sample tests are consistent in n (angles-based inference), so the probability of correctly calling X→Y tends to 1. Conversely, if no directed edge X→Y exists, all back-door and collider paths are blocked by the conditioning set, implying symmetric (null) directional dependence; the rule refrains from a causal call with probability → 1.

Note that when *M* is conditional on other variables, an edge X→Y indicates a direct causal relationship, but when *M* is marginal, the same edge indicates only a causal pathway or reachability, not necessarily a direct connection. In the empirical study conducted herein, conditioning was used, allowing for the former interpretation.

Finally, note again that ADD, as defined herein, identifies causal structure, but does not estimate effect sizes or counterfactual outcomes.

**B.** Below we place ADD within the Causal Inference Literature

**<u>Proposition</u>**:

Let $V = \{X_1, \ldots, X_p\}$ be generated by acyclic structural equations $X_j = f_j\left(X_{Pa(j)}, U_j\right)$ with mutually independent exogenous noises $U_j$. Under the causal Markov property and (qualified) faithfulness, the observed distribution is Markov to the causal DAG "G." We adopt these semantics and make the following explicit: (i) acyclicity; (ii) causal sufficiency (or an ancestral interpretation if violated); (iii) i.i.d./weak-stationarity; (iv) measurement reliability. Directional inference in ADD uses a directional dependence functional *M*; edges are interpreted as direct parents when *M* is evaluated conditionally on V\{X,Y}, and as directed reachability when evaluated marginally.

This makes ADD comparable with established approaches in the causal literature (see Zanga et al., 2025). Some of these include constraint-based methods, such as PC algorithms (see Kalisch, M., & Bühlmann, P. (2007)), FCI algorithms (see Sprites (2001)), and Markov Blanket algorithms (see Aliferis et al. (2010), and Chen et al., (2026)); they also include score-based methods, such as NOTEARS (see Zheng et al. (2018)) and its variants (e.g. DYNOTEARS, see Pamfil et al. (2020)), Greedy Equivalence Search (see Chickering, D., (2002) and Hauser and Bühlmann (2012)) and Kernel-based Score Functions (see Huang et al., (2018) and Wang et al., (2024)); finally, they include functional causal models, such as LiNGAM (see Shimizu et al. (2006)), and its variants such as VarLiNGAM (see Hyvärinen et al. (2010)). Unlike those examples, however, ADD takes a completely original approach by performing finite-sample inference with positive-definite constraints at the matrix level, which may very well improve coherence of edge calls across the graph, especially compared to those algorithms that identify causal edges sequentially. This will be tested in a more extensive, follow-up empirical study pitting ADD against many of the above-mentioned competitors.

**7. Appendix D: Conditioning Algorithm**

We condition on the factors whose edges are identified by ADD to make inferences regarding direct adjacency rather than reachability.

The conditioning algorithm is described below, step by step.

The conditioning approach is a residual-based approximation and conservative in that it conditions separately on both the parent and the child factors of an edge/cell, regardless of the direction of the dependence of the other factors on either. In other words, regressions are run separately on the parent and the child, using only those factors significantly associated with BOTH, in either direction, and the residuals of each regression are used to (re)assess the (potential) dependence between the parent and the child (note that the significance threshold used to identify regression covariates is, conservatively, the same Bonferroni-adjusted threshold as that used to identify significant edges). To avoid overfitting and instability, the model uses a simple linear OLS based on the two most significant factors of the remaining eight (out of ten) factors in the Digitale et al. (2022) DAG (if there are more than two significantly associated with BOTH the parent and the child). Higher order polynomial regressions undoubtedly would identify more cases that invite conditioning, but increasing power with regressions based on more, and more complex covariates would come at the price of increasing type 1 error, i.e. the approach would be more vulnerable to overfitting and instability. Conditioning on both directions, and separately on both the parent and child factors, while using straightforward models to do so provides a reasonable balance for this tradeoff, and a principled approach to conditioning in this setting.

Notably, just as with pruning, this conditioning induces non-positive definiteness in zero cases amongst all the 10,000 simulations run in the study. This result is not surprising vis-à-vis pruning, as our method of pruning places the all-pairwise dependence matrix deeper into the convex cone and thus, farther from singularity. Conditioning appears to be having a similar effect here, as it generally will dampen observed associations and cause ADD to miss some true edges even as it is correctly rejecting more nonexistent edges. Nevertheless, positive definiteness always is verified empirically, and even though not encountered under these data conditions, any singular edge cases can be made positive definite using "nearest PD matrix" algorithms like Higham (2002). This approach, however, would not be advisable in situations where cases of non-positive definiteness occur systematically, say, more than one percent of the time, as this would distort the sample space of the dependence measure matrix, invalidate associated inference, and call into question the wisdom of using that particular dependence measure at all, at least under the given data conditions. But that is not an issue herein.

Conditioning Algorithm:

Nsims = 10000;
For each of Nsims simulations, DO:

1. Calculate two all-pairwise dependence measure matrices (for example, using improved Chatterjee's correlation), one in each direction
2. Identify cells (factor pairs) from 1. with a statistically significant dependence measure value (where $\alpha$=0.025, as described above in the text), in EITHER direction
3. For each cell/pair in 2., identify those remaining factors that are statistically significantly related to BOTH variables (parent and child) of the pair, regardless of the direction of association (note that the same Bonferroni-adjusted significance threshold ($\alpha$=0.025) is used here to conservatively identify relevant conditioning factors)
4. If there are more than two factors identified in 3. for a given pair/cell, keep only the two largest, where "largest" is defined as follows: each factor has two directional effects on each parent and each child – take the maximum of the two for the parent, and the maximum of the two for the child; then add these two (maximum) effects for each factor, rank these sums across the remaining factors and keep only the 2 largest.
5. Run two OLS regressions using the factors identified in 3. and 4.: one regression on the parent and one regression on the child, and keep the residuals of both regressions
6. (Re)calculate the dependence measure between the parent and child, in both directions, based on the residuals obtained from 5. and insert this value into both the all-pairwise dependence measure matrices, one in each direction (and insert the value in both cell i,j and cell j,i as the matrix must remain symmetric, even though the dependence measure is asymmetric). Verify that the matrix remains positive definite (apply Higham (2002) if rare edge cases are singular).
7. Apply pruning as described in the text.

END;